\documentclass[10pt]{article}
\pdfoutput=1
\usepackage{amsmath}
\usepackage{amssymb}
\usepackage{graphicx}

\title{A wave equation including leptons and quarks for the standard model of quantum physics in Clifford Algebra}
\author{
Claude Daviau\\
Le Moulin de la Lande\\
44522 Pouill\'e-les-coteaux\\
France\\
email: claude.daviau@nordnet.fr\\ \\
Jacques Bertrand\\
15 avenue Danielle Casanova\\
95210 Saint-Gratien\\
France\\
email: bertrandjacques-m@orange.fr 
}

\begin{document}

\maketitle

\begin{abstract}
A wave equation with mass term is studied for all particles and antiparticles of the first 
generation: electron and its neutrino, positron and antineutrino, quarks $u$ and $d$ with 
three states of color and antiquarks $\overline{u}$ and $\overline{d}$. This wave equation 
is form invariant under the $Cl_3^*$ group generalizing the relativistic invariance. It is 
gauge invariant under the $U(1)\times SU(2) \times SU(3)$ group of the standard model of 
quantum physics. The wave is a function of space and time with value in the Clifford algebra 
$Cl_{1,5}$. All features of the standard model, charge conjugation, color, left waves, 
Lagrangian formalism, are linked to the geometry of this extended space-time. 
\end{abstract}
\section*{Introduction}

We use here all notations of ``New insights in the standard model of quantum physics in Clifford Algebra" 
\cite{davi:14}. The wave equation for all particles of the first generation is a generalization of 
the wave equation obtained in 6.7 for the electron and its neutrino. This wave equation has obtained
a proper mass term compatible with the gauge invariance in \cite{dabe:14}. It is a generalization of the
homogeneous nonlinear Dirac equation for the electron alone \cite{davi:93} \cite{davi:97} \cite{davi:05}
\cite{davi:11} \cite{dav2:12} \cite{davi:12} \cite{dav:12}. The link with the usual presentation of
the standard model is made by the left and right Weyl spinors used for waves of each particle.
These right and left waves are parts of the wave with value in $Cl_{1,5}$. 

We used previously the same algebra $Cl_{5,1}=Cl_{1,5}$. It is the same algebra, and this explains very
well why sub-algebras $Cl_{1,3}$ and $Cl_{3,1}$ have been equally used to describe relativistic
physics \cite{dehe:93} \cite{hest:86}. But the signature of the scalar product cannot be free,
this scalar product being linked to the gravitation in the general relativity. It happens that
vectors of $Cl_{1,5}$ are pseudo-vectors of $Cl_{5,1}$ and more generally that $n$-vectors of
$Cl_{1,5}$ are $(6-n)$-vectors of $Cl_{5,1}$. The generalization of the wave equation for
electron-neutrino is simpler if we use $Cl_{1,5}$. This is the first indication that the
signature $+-----$ is the true one. We explain in Appendix A how the reverse in $Cl_{1,5}$
is linked to the reverse in $Cl_{1,3}$, a necessary condition to get the wave equation of
all particles of the first generation. 

We have noticed, for the electron alone firstly (See \cite{davi:12} 2.4), next for
electron+neutrino \cite{dabe:14} the double link existing between the wave equation
and the Lagrangian density: It is well known that the wave equation may be obtained 
from the Lagrangian density by the variational calculus. The new link is that the
real part of the invariant wave equation is simply $\mathcal{L}=0$. The Lagrangian formalism
is then necessary, being a consequence of the wave equation. Next we have extended
the double link to electro-weak interactions in the leptonic case (electron + neutrino).
Now we are extending the double link to the gauge group of the standard model. The
Lagrangian density must then be the real part of the invariant wave equation.

Moreover we generalized the non-linear homogeneous wave equation of the electron, and
we got a wave equation with mass term \cite{dabe:14}, form invariant under the
$Cl_3^*=GL(2,\mathbb{C})$ group and gauge invariant under the $U(1)\times SU(2)$
gauge group of electro-weak interactions. Our aim is to explain how this may be
extended to a wave equation with mass term, both form invariant under $Cl_3^*$
and gauge invariant under the $U(1)\times SU(2)\times SU(3)$ gauge group of the
standard model, including both electro-weak and strong interactions.

\section{From the lepton case to the full wave}\setcounter{equation}{0}

The standard model adds to the leptons (electron $e$ and its neutrino $n$) in the ``first generation"
two quarks $u$ and $d$ with three states each. Weak interactions acting only on left waves of
quarks (and right waves of antiquarks) we read the wave of all fermions of the first generation
as follows:
\begin{align}
\Psi &=\begin{pmatrix} \Psi_l & \Psi_r \\ \Psi_g & \Psi_b \end{pmatrix}~;~~\Psi_l=
\begin{pmatrix} \phi_{e} & \phi_{n} \\ \widehat{\phi}_{n} & \widehat{\phi}_{e} \end{pmatrix}
=\begin{pmatrix} \phi_{e} & \phi_{n} \\ \widehat{\phi}_{\overline n}\sigma_1
 & \widehat{\phi}_{\overline e}\sigma_1 \end{pmatrix} \\ \Psi_r&=
\begin{pmatrix} \phi_{dr} & \phi_{ur} \\ \widehat{\phi}_{ur} & \widehat{\phi}_{dr} \end{pmatrix}
=\begin{pmatrix} \phi_{dr} & \phi_{ur} \\ \widehat{\phi}_{\overline ur}\sigma_1
 & \widehat{\phi}_{\overline dr}\sigma_1 \end{pmatrix};~\Psi_g=
\begin{pmatrix} \phi_{dg} & \phi_{ug} \\ \widehat{\phi}_{ug} & \widehat{\phi}_{dg} \end{pmatrix}
=\begin{pmatrix} \phi_{dg} & \phi_{ug} \\ \widehat{\phi}_{\overline ug}\sigma_1
 & \widehat{\phi}_{\overline dg}\sigma_1 \end{pmatrix} \notag\\ \Psi_b&=
\begin{pmatrix} \phi_{db} & \phi_{ub} \\ \widehat{\phi}_{ub} & \widehat{\phi}_{db} \end{pmatrix}
=\begin{pmatrix} \phi_{db} & \phi_{ub} \\ \widehat{\phi}_{\overline ub}\sigma_1
 & \widehat{\phi}_{\overline db}\sigma_1 \end{pmatrix}
\end{align}
The electro-weak theory \cite{wein:67} needs three spinorial waves in the electron-neutrino case:
the right $\xi_e$ and the left $\eta_e$ of the electron and the left spinor $\eta_n$ of the electronic
neutrino. The form invariance of the Dirac theory imposes to use $\phi_e$ for the electron and
$\phi_n$ for its neutrino satisfying
\begin{align}
\phi_e &=\sqrt{2}\begin{pmatrix}
\xi_{1e} & -\eta_{2e}^* \\ \xi_{2e} &\eta_{1e}^*
\end{pmatrix}=\sqrt{2}(\xi_e~~-i\sigma_2 \eta_e^*)~;~~\phi_n=\sqrt{2}\begin{pmatrix}
0 & -\eta_{2n}^* \\0 &\eta_{1n}^* \end{pmatrix}\\
\widehat\phi_e &=\sqrt{2}\begin{pmatrix}
\eta_{1e} & -\xi_{2e}^* \\ \eta_{2e} &\xi_{1e}^*
\end{pmatrix}=\sqrt{2}(\eta_e~~-i\sigma_2\xi_e^*)~;~~\widehat\phi_n=\sqrt{2}\begin{pmatrix}
\eta_{1n} & 0 \\ \eta_{2n}&0 \end{pmatrix}=\sqrt{2}(\eta_n~~0).
\end{align}
Waves $\phi_e$ and $\phi_n$ are functions of space and time with value into the Clifford algebra $Cl_3$
of the physical space. The standard model uses only a left $\eta_n$ wave for the neutrino. 
We always use the matrix representation (A.1) which allows to see the Clifford algebra $Cl_{1,3}$ 
as a sub-algebra of $M_4(\mathbb{C})$. Under the dilation $R$ with ratio $r$ induced by any $M$ in
$GL(2,\mathbb{C})$ we have (for more details, see \cite{davi:11}):
\begin{align}
x'&=M x M^\dagger~;~~\det(M)=r e^{i\theta}~;~~x=x^\mu\sigma_\mu~;~~x'={x'}^\mu\sigma_\mu \\
\xi '&= M \xi~;~~\eta '=\widehat{M}\eta~;~~\eta_n '=\widehat{M}\eta_n~;~~\phi_e ' = M\phi_e
~;~~\phi_n '=M \phi_n \\ \Psi_l ' &=\begin{pmatrix}
\phi_e' & \phi_n '\\ \widehat{\phi}_n' &\widehat{\phi}_e'
\end{pmatrix}=\begin{pmatrix}
M & 0 \\0 & \widehat{M} \end{pmatrix}\begin{pmatrix}
\phi_e & \phi_n \\ \widehat{\phi}_n &\widehat{\phi}_e
\end{pmatrix}=N \Psi_l
\end{align}
The form (1.3) of the wave is compatible both with the form invariance of the Dirac theory and with
the charge conjugation used in the standard model: the wave $\psi_{\overline{e}}$ of the positron satisfies
\begin{equation}
\psi_{\overline{e}}=i\gamma_2\psi^*\Leftrightarrow\widehat{\phi}_{\overline{e}}=\widehat{\phi}_e\sigma_1
\end{equation}
We can then think the $\Psi_l$ wave as containing the electron wave $\phi_e$, the neutrino wave
$\phi_n$ and also the positron wave $\phi_{\overline{e}}$ and the antineutrino wave $\phi_{\overline{n}}$:
\begin{equation}
\Psi_l=\begin{pmatrix}
\phi_e & \phi_n \\ \widehat{\phi}_{\overline{n}}\sigma_1 &\widehat{\phi}_{\overline{e}}\sigma_1
\end{pmatrix}~;~~\phi_{\overline{e}} =\sqrt{2}\begin{pmatrix}
\xi_{1\overline{e}} & -\eta_{2\overline{e}}^* \\ \xi_{2\overline{e}} &\eta_{1\overline{e}}^*
\end{pmatrix}~;~~\phi_{\overline{n}}=\sqrt{2}\begin{pmatrix}
\xi_{1\overline{n}} & 0 \\\xi_{2\overline{n}} & 0  \end{pmatrix}
\end{equation}
And the antineutrino has only a right wave. The multivector $\Psi_l(x)$ is usually an invertible element
of the space-time algebra because (See \cite{davi:14} (6.250)) with:
\begin{align}
a_1&=\det(\phi_e)=\phi_e \overline{\phi}_e=2(\xi_{1e}\eta_{1e}^*+\xi_{2e}\eta_{2e}^*) \\
a_2&=2(\xi_{1\overline{e}}\eta_{1n}^*+\xi_{2\overline{e}}\eta_{2n}^*)=2(\eta_{2e}^*\eta_{1n}^*-
\eta_{1e}^*\eta_{2n}^*)\\
a_3&=2(\xi_{1e}\eta_{1n}^*+\xi_{2e}\eta_{2n}^*)
\end{align}
we got
\begin{align}
\det(\Psi_l)&=a_1 a_1^*+a_2 a_2^*.
\end{align}
Most of the preceding presentation is easily extended to quarks. For each color $c=r,g,b$ the
electro-weak theory needs only left waves:
\begin{equation}
\Psi_c=\begin{pmatrix}
\phi_{dc}&\phi_{uc}\\ \widehat{\phi}_{uc}& \widehat{\phi}_{dc}
\end{pmatrix}
;~\widehat{\phi}_{dc}=\sqrt{2}\begin{pmatrix}
\eta_{1dc}&0\\ \eta_{2dc}&0
\end{pmatrix}
;~\widehat{\phi}_{uc}=\sqrt{2}\begin{pmatrix}
\eta_{1uc}&0\\ \eta_{2uc}&0
\end{pmatrix}
\end{equation}
The $\Psi$ wave is now a function of space and time with value into $Cl_{1,5}=Cl_{5,1}$ which is a
sub-algebra (on the real field) of $Cl_{5,2}=M_8(\mathbb{C})$:
\begin{equation}
\Psi=\begin{pmatrix}
\Psi_l &\Psi_r \\ \Psi_g & \Psi_b
\end{pmatrix};~\widetilde{\Psi}=\begin{pmatrix}
\widetilde\Psi_b & \widetilde\Psi_r \\ \widetilde\Psi_g &\widetilde\Psi_l
\end{pmatrix}
\end{equation}
The link between the reverse in $Cl_{1,5}$ and the reverse in $Cl_{1,3}$ is not trivial and is
detailed in Appendix A. The wave equation for all objects of the first generation reads
\begin{equation}
0=(\underline{\mathbf{D}}\Psi)L_{012}+\underline{\mathbf{M}}
\end{equation}
The mass term reads
\begin{equation}
\underline{\mathbf{M}}=\begin{pmatrix}
m_2 \rho_2 \chi_b & m_2 \rho_2 \chi_g \\ m_2 \rho_2\chi_r & m_1 \rho_1 \chi_l
\end{pmatrix}
\end{equation}
where we use the scalar densities $s_j$ and $\chi$ terms of Appendix B, with
\begin{equation}
\rho_1^2=a_1 a_1^*+a_2 a_2^*+a_3 a_3^*~;~~\rho_2^2=\sum_{j=1}^{j=15}s_j s_j^*.
\end{equation}
The covariant derivative $\underline{\mathbf{D}}$ uses the
matrix representation (A.1) and reads
\begin{align}
\underline{\mathbf{D}}&=\underline{\pmb\partial} +\frac{g_1}{2}\underline{\mathbf{B}}\ \underline{P}_0
+\frac{g_2}{2}\underline{\mathbf{W}}^j \underline{P}_j +\frac{g_3}{2} \underline{\mathbf{G}}^k \underline{\mathbf{i}}\Gamma_k \\
\underline{\mathbf{D}}&=\sum_{\mu=0}^{3}L^\mu D_\mu~;~~\underline{\pmb\partial}
=\sum_{\mu=0}^{3}L^\mu \partial_\mu~;~~\underline{\mathbf{B}}
=\sum_{\mu=0}^{3}L^\mu  B_\mu\\
\underline{\mathbf{W}}^j&=\sum_{\mu=0}^{3}L^\mu  W_\mu^j,~j=1,2,3\\
\underline{\mathbf{G}}^k&=\sum_{\mu=0}^{3}L^\mu  G_\mu^k,~k=1,2,\dots, 8.
\end{align}
We use two projectors  satisfying
\begin{equation}
\underline{P}_\pm(\Psi)=\frac{1}{2}(\Psi\pm \underline{\mathbf{i}}\Psi L_{21})~;~~
\underline{\mathbf{i}}=L_{0123}
\end{equation}
Three operators act on quarks like on leptons:
\begin{align}
\underline{P}_1(\Psi)&=\underline{P}_+(\Psi)L_{35}\\
\underline{P}_2(\Psi)&=\underline{P}_+(\Psi)L_{5012}\\
\underline{P}_3(\Psi)&=\underline{P}_+(\Psi)(-\underline{\mathbf{i}}).
\end{align}
The fourth operator acts differently on the leptonic and on the quark sector.
Using projectors:
\begin{equation}
P^+=\frac{1}{2}(I_8+L_{012345})=\begin{pmatrix}
I_4 & 0 \\ 0&0\end{pmatrix};~P^-=\frac{1}{2}(I_8-L_{012345})=\begin{pmatrix}
0 & 0 \\ 0&I_4\end{pmatrix}
\end{equation}
we can separate the lepton part $\Psi^l$ and the quark part $\Psi^c$ of
the wave:
\begin{equation}
\Psi^l=P^+\Psi P^+=\begin{pmatrix}\Psi_l & 0 \\ 0&0
\end{pmatrix} ;~\Psi^c=\Psi-\Psi^l=\begin{pmatrix}
0 & \Psi_r \\ \Psi_g & \Psi_b
\end{pmatrix}.
\end{equation}
and we get (see \cite{davi:14} (B.4) with $a=b=1$)
\begin{align}
\underline{P}_0(\Psi^l)&=\frac{1}{2}\underline{\mathbf{i}}\Psi^l L_{21}+\frac{3}{2}\Psi^l L_{21} \\
\underline{P}_0(\Psi^c)&=-\frac{1}{3}\Psi^c L_{21}.
\end{align}
This last relation comes from the non-existence of the right part of the $\Psi^c$ waves.

\section{Chromodynamics}\setcounter{equation}{0}

We start from generators $\lambda_k$ of the $SU(3)$ gauge group of chromodynamics
\begin{align}
\lambda_1 &=\begin{pmatrix}0&1&0 \\1&0&0 \\0&0&0 \end{pmatrix},~
\lambda_2 =\begin{pmatrix}0&-i&0 \\i&0&0 \\0&0&0 \end{pmatrix},~
\lambda_3 =\begin{pmatrix}1&0&0 \\0&-1&0 \\0&0&0 \end{pmatrix}\notag \\
\lambda_4 &=\begin{pmatrix}0&0&1 \\0&0&0 \\1&0&0 \end{pmatrix},~
\lambda_5 =\begin{pmatrix}0&0&-i \\0&0&0 \\i&0&0 \end{pmatrix},~
\lambda_6 =\begin{pmatrix}0&0&0 \\0&0&1 \\0&1&0 \end{pmatrix}\notag \\
\lambda_7 &=\begin{pmatrix}0&0&0 \\0&0&-i \\0&i&0 \end{pmatrix},~
\lambda_8 =\frac{1}{\sqrt{3}}\begin{pmatrix}1&0&0 \\0&1&0 \\0&0&-2 \end{pmatrix}.
\end{align}
To simplify here notations we use now $l$, $r$, $g$, $b$ instead $\Psi_l$, $\Psi_r$, $\Psi_g$, $\Psi_b$. 
So we  have $\Psi=\begin{pmatrix}l & r \\g & b \end{pmatrix}$.
Then (C.1) gives
\begin{align}
\lambda_1 \begin{pmatrix}r\\g\\b \end{pmatrix}&=\begin{pmatrix}g\\r\\0 \end{pmatrix},~
\lambda_2 \begin{pmatrix}r\\g\\b \end{pmatrix}=\begin{pmatrix}-ig\\ir\\0 \end{pmatrix},~
\lambda_3 \begin{pmatrix}r\\g\\b \end{pmatrix}=\begin{pmatrix}r\\-g\\0 \end{pmatrix}\notag\\
\lambda_4 \begin{pmatrix}r\\g\\b \end{pmatrix}&=\begin{pmatrix}b\\0\\r \end{pmatrix},~
\lambda_5 \begin{pmatrix}r\\g\\b \end{pmatrix}=\begin{pmatrix}-ib\\0\\ir \end{pmatrix},~
\lambda_6 \begin{pmatrix}r\\g\\b \end{pmatrix}=\begin{pmatrix}0\\b\\g \end{pmatrix}\\
\lambda_7 \begin{pmatrix}r\\g\\b \end{pmatrix}&=\begin{pmatrix}0\\-ib\\ig \end{pmatrix},~
\lambda_8 \begin{pmatrix}r\\g\\b \end{pmatrix}=\frac{1}{\sqrt{3}}\begin{pmatrix}r\\g\\-2b \end{pmatrix}.\notag
\end{align}
We name $\Gamma_k$ operators corresponding to $\lambda_k$ acting on $\Psi$. We get with projectors
$P^+$ and $P^-$ in (1.27):
\begin{align}
\Gamma_1 (\Psi)&=\frac{1}{2}(L_4\Psi L_4+L_{01235}\Psi L_{01235})=
\begin{pmatrix}0 &g\\r &0 \end{pmatrix}\\
\Gamma_2 (\Psi)&=\frac{1}{2}(L_5\Psi L_4 -L_{01234}\Psi L_{01235})=
\begin{pmatrix}0 &-\mathbf{i}g\\\mathbf{i}r &0 \end{pmatrix} \\
\Gamma_3 (\Psi)&=P^+\Psi P^- - P^- \Psi P^+ =
\begin{pmatrix}0 &r\\-g &0 \end{pmatrix}\\
\Gamma_4 (\Psi)&=L_{01253}\Psi P^- = \begin{pmatrix}0 &b\\0 &r \end{pmatrix}~;~
\Gamma_5(\Psi)=L_{01234} \Psi P^-
=\begin{pmatrix}0 &-\mathbf{i}b\\0 &\mathbf{i}r\end{pmatrix}\\
\Gamma_6(\Psi)&= P^- \Psi L_{01253}=\begin{pmatrix}0 &0\\b &g\end{pmatrix}~;~
\Gamma_7(\Psi)=-\underline{\mathbf{i}}P^- \Psi L_4=\begin{pmatrix}0 &0\\-\mathbf{i}b &\mathbf{i}g
\end{pmatrix}\\
\Gamma_8(\Psi)&=\frac{1}{\sqrt{3}}(P^-\Psi L_{012345}+L_{012345}\Psi P^-)
=\frac{1}{\sqrt{3}}\begin{pmatrix}0 &r\\g &-2b\end{pmatrix}.
\end{align}
Everywhere the left up term is $0$, so all
$\Gamma_k$ project the wave $\Psi$ on its quark sector. 

We can extend the covariant derivative of electro-weak interactions in the electron-neutrino case:
\begin{equation}
D\Psi_l= \partial\Psi_l +\frac{g_1}{2} B P_0(\Psi_l)
+\frac{g_2}{2} W^j P_j(\Psi_l)
\end{equation}
to get the covariant derivative of the standard model
\begin{equation}
\underline D (\Psi)= \underline \partial(\Psi) +\frac{g_1}{2}\underline B \  \underline P_0(\Psi)
+\frac{g_2}{2}\underline W^j \underline P_j(\Psi)+\frac{g_3}{2}\underline G^k\underline{\mathbf{i}}
\Gamma_k(\Psi).
\end{equation}
where $g_3$ is another constant and $\underline G^k$ are eight terms called ``gluons''. Since $I_4$ commute
with any element of $Cl_{1,3}$ and since $P_j(\mathbf{i} \Psi_{ind})=\mathbf{i} P_j(\Psi_{ind})$
for $j=0,1,2,3$ and $ind=l,r,g,b$ each operator $\underline{\mathbf{i}} \Gamma_k$ commutes with
all operators $\underline P_j$.

Now we use 12 real numbers $a^0$, $a^j,~j=1,2,3$, $b^k,~k=1,2,...,8$, we let
\begin{equation}
S_0=a^0 \underline P_0;~S_1=\sum_{j=1}^{j=3} a^j \underline P_j;~S_2=\sum_{k=1}^{k=8}b^k
\underline{\mathbf{i}} \Gamma_k;~S=S_0 + S_1 +S_2
\end{equation}
and we get, using exponentiation 
\begin{align}
\exp(S) &=\exp(S_0)\exp(S_1)\exp(S_2)=\exp(S_1)\exp(S_0)\exp(S_2)\notag\\
&=\exp(S_0)\exp(S_2)\exp(S_1)=\dots
\end{align}
in any order. The set of these operators $\exp(S)$ is a $U(1)\times SU(2)\times SU(3)$ Lie group. 
Only difference with the standard model: the structure of this group is not postulated but calculated.
With 
\begin{equation}
\Psi '=[\exp(S)](\Psi)~;~\underline{\mathbf{D}}=L^\mu \underline D_\mu
~;~\underline{\mathbf{D}}'=L^\mu \underline D_\mu '
\end{equation}
the gauge transformation reads
\begin{align}
\underline D_\mu ' \Psi '&=[\exp(S)](\underline D_\mu \Psi)\\
B_\mu '&=B_\mu-\frac{2}{g_1}\partial_\mu a^0\\
{W '}_\mu^j \underline P_j&=\Big[\exp(S_1)W_\mu^j \underline P_j-\frac{2}{g_2}\partial_\mu[\exp(S_1)]
\Big] \exp(-S_1)\\
{\underline G '}_\mu^k \underline{\mathbf{i}}\Gamma_k&=\Big[\exp(S_2)\underline{G}_\mu^k 
\underline{\mathbf{i}}\Gamma_k-\frac{2}{g_3} \partial_\mu[\exp(S_2)] \Big] \exp(-S_2).
\end{align}
The $SU(3)$ group generated by projectors $\Gamma_k$ acts only on the quark sector of the wave:
\begin{equation}
P^+[\exp(b^k\underline{\mathbf{i}} \Gamma_k](\Psi)P^+= 
P^+\Psi  P^+ = \Psi^l
\end{equation}
The physical translation is: Leptons do not act by strong interactions. This comes from the structure of the
wave itself. It is fully satisfied in experiments.
We get then a $U(1)\times SU(2)\times SU(3)$ gauge group for a wave including all fermions of
the first generation. This group acts on the lepton sector only by its $U(1)\times SU(2)$
part. Consequently the wave equation is composed of a lepton wave equation and a quark
wave equation:
\begin{align}
0&=(\underline{\mathbf{D}}\Psi^l)L_{012}+m_1 \rho_1\begin{pmatrix}
0 & 0 \\ 0 & \chi_l
\end{pmatrix}\\
0&=(\underline{\mathbf{D}}\Psi^c)L_{012}+m_2 \rho_2\chi^c;~\chi^c=\begin{pmatrix}
\chi_b & \chi_g \\ \chi_r & 0
\end{pmatrix}
\end{align}
The wave equation (2.19) is equivalent to the wave equation
\begin{equation}
\mathbf{D}\Psi_l\gamma_{012}+m_1\rho_1 \chi_l=0~;~~\gamma_{012}=\gamma_0\gamma_1\gamma_2
\end{equation}
studied in \cite{dabe:14} \cite{dav:14}, where
\begin{align}
\chi_l &=\frac{1}{\rho_1^2}\begin{pmatrix}
a_1^*\phi_e+a_2^*\phi_n\sigma_1+a_3^*\phi_n & -a_2^*\phi_{eL}\sigma_1+a_3^*
\phi_{eR} \\ a_2\widehat{\phi}_{eL}\sigma_1+a_3\widehat{\phi}_{eR} &
a_1\widehat{\phi}_e -a_2\widehat{\phi}_n \sigma_1 + a_3\widehat{\phi}_n
\end{pmatrix}\\
\phi_{eR}&=\phi_e\frac{1+\sigma_3}{2}~;~~\phi_{eL}=\phi_e\frac{1-\sigma_3}{2}
\end{align}
This wave equation is equivalent to the invariant equation:
\begin{equation}
\widetilde \Psi_l (\mathbf{D}\Psi_l)\gamma_{012}+m\rho_1 \widetilde \Psi_l\chi_l=0~;~~
\widetilde \Psi_l=\begin{pmatrix}
\overline{\phi}_e &\phi_n^\dagger \\ \overline{\phi}_n & \phi_e^\dagger
\end{pmatrix}.
\end{equation}
This wave equation is form invariant under the Lorentz dilation $R$ induced by any
invertible matrix $M$ satisfying (1.5), (1.6), (1.7) \cite{davi:14}. It is gauge invariant 
under the $U(1)\times SU(2)$ group \cite{dabe:14} generated by operators $P_\mu$ which are projections 
on the lepton sector of the operators defined in (1.23) to (1.29). 
Therefore we need to study only the quark sector and its wave equation (2.20).

We begin by the double link between
wave equation and Lagrangian density that we have remarked firstly in the Dirac equation
\cite{davi:12}, next in the lepton case electron+neutrino \cite{davi:14}.

\section{Double link between wave equation and Lagrangian density}\setcounter{equation}{0}

The existence of a Lagrangian mechanism in optics and mechanics is known since Fermat
and Maupertuis. This principle of minimum is everywhere in quantum mechanics from its
beginning, it is the main reason of the hypothesis of a wave linked to the move of any
material particle made by L. de Broglie \cite{brog:24}. By the calculus of variations it is
always possible to get the wave equation from the Lagrangian density. But another link
exists : the Lagrangian density is the real scalar part of the invariant wave equation.
This was obtained firstly for the electron alone \cite{davi:12}, next for the pair
electron-neutrino \cite{dabe:14} where the Lagrangian density reads
\begin{align}
\mathcal{L}_l&=\mathcal{L}_0+g_1 \mathcal{L}_1+ g_2 \mathcal{L}_2 +m_1\rho_1\\
\mathcal{L}_0&=\Re[-i(\eta_e^\dagger \sigma^\mu \partial_\mu \eta_e + \xi_e^\dagger
\widehat{\sigma}^\mu \partial_\mu \xi_e +\eta_n^\dagger \sigma^\mu \partial_\mu \eta_n)]\\
\mathcal{L}_1&=B_\mu(\frac{1}{2}\eta_e^\dagger \sigma^\mu \eta_e + \xi_e^\dagger\widehat{\sigma}^\mu \xi_e
+\frac{1}{2}\eta_n^\dagger \sigma^\mu \eta_n )\\
\mathcal{L}_2&=-\Re[(W_\mu^1+iW_\mu^2)\eta_e^\dagger\sigma^\mu \eta_n]+\frac{W_\mu^3}{2}
(\eta_e^\dagger \sigma^\mu \eta_e - \eta_n^\dagger \sigma^\mu \eta_n).
\end{align}
We shall establish the double link now for the wave equation (1.16). It is sufficient to add the property
for (2.20). This equation is equivalent to the invariant equation:
\begin{align}
0&=\widetilde{\Psi}^c(\underline{\mathbf{D}}\Psi^c)L_{012}+m_2 \rho_2\widetilde{\Psi}^c
\chi^c\\ \widetilde{\Psi}^c&=\begin{pmatrix}
\widetilde{\Psi}_b & \widetilde{\Psi}_r \\ \widetilde{\Psi}_g &0
\end{pmatrix};~\chi^c= \begin{pmatrix}
\chi_b & \chi_g \\ \chi_r & 0
\end{pmatrix}
\end{align}
We get from the covariant derivative (1.19) with the operators $\underline{P}_j$ in (1.24), 
(1.25), (1.26) and (1.30) and $\Gamma_k$ in (2.3) to (2.8) and with $\Psi^c$ in (1.28)
\begin{align}
\underline{\mathbf{D}}\Psi^c&=\begin{pmatrix}
A_g & A_b \\0 & A_r \end{pmatrix} \\
A_g &= \pmb\partial \Psi_g-\frac{g_1}{6}\mathbf{B} \Psi_g \gamma_{21}+\frac{g_2}{2}
(\mathbf{W}^1 \Psi_g \gamma_3\mathbf{i}+\mathbf{W}^2 \Psi_g \gamma_3-
\mathbf{W}^3 \Psi_g \mathbf{i})\notag \\
&+\frac{g_3}{2}(\mathbf{G}^1 \mathbf{i}\Psi_r-\mathbf{G}^2 \Psi_r
-\mathbf{G}^3 \mathbf{i}\Psi_g+\mathbf{G}^6 \mathbf{i}\Psi_b+\mathbf{G}^7 \Psi_b
+\frac{1}{\sqrt{3}}\mathbf{G}^8 \mathbf{i}\Psi_g)
\end{align}
\begin{align}
A_b &= \pmb\partial \Psi_b-\frac{g_1}{6}\mathbf{B} \Psi_b \gamma_{21}+\frac{g_2}{2}
(\mathbf{W}^1 \Psi_b \gamma_3\mathbf{i}+\mathbf{W}^2 \Psi_b \gamma_3-
\mathbf{W}^3 \Psi_b \mathbf{i})\notag \\
&+\frac{g_3}{2}(\mathbf{G}^4 \mathbf{i}\Psi_r-\mathbf{G}^5 \Psi_r
+\mathbf{G}^6 \mathbf{i}\Psi_g-\mathbf{G}^7 \Psi_g
-\frac{2}{\sqrt{3}}\mathbf{G}^8 \mathbf{i}\Psi_b)\\
A_r &= \pmb\partial \Psi_r-\frac{g_1}{6}\mathbf{B} \Psi_r \gamma_{21}+\frac{g_2}{2}
(\mathbf{W}^1 \Psi_r \gamma_3\mathbf{i}+\mathbf{W}^2 \Psi_r \gamma_3-
\mathbf{W}^3 \Psi_r \mathbf{i})\notag \\
&+\frac{g_3}{2}(\mathbf{G}^1 \mathbf{i}\Psi_g+\mathbf{G}^2 \Psi_g
+\mathbf{G}^3 \mathbf{i}\Psi_r+\mathbf{G}^4 \mathbf{i}\Psi_b+\mathbf{G}^5 \Psi_b
+\frac{1}{\sqrt{3}}\mathbf{G}^8 \mathbf{i}\Psi_r)
\end{align}
Next we get
\begin{align}
&\widetilde{\Psi}^c(\underline{\mathbf{D}}\Psi^c)L_{012}+m_2 \rho_2\widetilde{\Psi}^c
\chi^c\\ &=\begin{pmatrix}
\widetilde{\Psi}_b(A_b \gamma_{012}+m_2\rho_2\chi_b)+\widetilde{\Psi}_r(A_r \gamma_{012}+m_2\rho_2\chi_r)
& \widetilde{\Psi}_b(A_g \gamma_{012}+m_2\rho_2\chi_g)\\ \widetilde{\Psi}_g(A_b \gamma_{012}+m_2\rho_2\chi_b)
&\widetilde{\Psi}_g(A_g \gamma_{012}+m_2\rho_2\chi_g)
\end{pmatrix}\notag
\end{align}
The calculation of the Lagrangian density in the general case is similar to the
lepton case. We get
\begin{align}
\mathcal{L}&=\mathcal{L}_l+ \mathcal{L}_c \\
\mathcal{L}_c&=\sum_{c=r,g,b}\mathcal{L}_{0c}+g_1\sum_{c=r,g,b}\mathcal{L}_{1c}
+g_2\sum_{c=r,g,b}\mathcal{L}_{2c}+g_3\mathcal{L}_3+m_2\rho_2
\end{align}
The calculation of $\mathcal{L}_{jc},~j=0,1,2$ replaces the pair e-n by the pair dc-uc and
suppress the $\xi$ terms, then (3.2) (3.3) (3.4) become 
\begin{align}
\mathcal{L}_{0c}&=\Re[-i(\eta_{dc}^\dagger \sigma^\mu \partial_\mu \eta_{dc}
+\eta_{uc}^\dagger \sigma^\mu \partial_\mu \eta_{uc})]\\
\mathcal{L}_{1c}&=-\frac{B_\mu}{6}(\eta_{dc}^\dagger \sigma^\mu \eta_{dc}
+\eta_{uc}^\dagger \sigma^\mu \eta_{uc})\\
\mathcal{L}_{2c}&=-\Re[(W_\mu^1+i W_\mu^2)\eta_{dc}^\dagger\sigma^\mu \eta_{uc}]
+\frac{W_\mu^3}{2}(\eta_{dc}^\dagger\sigma^\mu \eta_{dc}-\eta_{uc}^\dagger\sigma^\mu \eta_{uc})
\end{align}
Since three $SU(2)$ group are included in $SU(3)$ the calculation of $\mathcal{L}_3$
has similarities with the calculation of $\mathcal{L}_2$ and we get
\begin{align}
\mathcal{L}_3=&-\Re[(G_\mu^1+iG_\mu^2)(\eta_{dr}^\dagger\sigma^\mu \eta_{dg}
+\eta_{ur}^\dagger\sigma^\mu \eta_{ug})]\\
&-\Re[(G_\mu^4+iG_\mu^5)(\eta_{dr}^\dagger\sigma^\mu \eta_{db}
+\eta_{ur}^\dagger\sigma^\mu \eta_{ub})]\notag \\
&-\Re[(G_\mu^6+iG_\mu^7)(\eta_{dg}^\dagger\sigma^\mu \eta_{db}
+\eta_{ug}^\dagger\sigma^\mu \eta_{ub})]\notag \\
&+\frac{G_\mu^3}{2}(-\eta_{dr}^\dagger\sigma^\mu \eta_{dr}
-\eta_{ur}^\dagger\sigma^\mu \eta_{ur}+\eta_{dg}^\dagger\sigma^\mu \eta_{dg}
+\eta_{ug}^\dagger\sigma^\mu \eta_{ug})\notag\\
&+\frac{G_\mu^8}{2\sqrt{3}}(-\eta_{dr}^\dagger\sigma^\mu \eta_{dr}
-\eta_{ur}^\dagger\sigma^\mu \eta_{ur}+\eta_{db}^\dagger\sigma^\mu \eta_{db}
+\eta_{ub}^\dagger\sigma^\mu \eta_{ub})\notag\\
&+\frac{G_\mu^8}{2\sqrt{3}}(-\eta_{dg}^\dagger\sigma^\mu \eta_{dg}
-\eta_{ug}^\dagger\sigma^\mu \eta_{ug}+\eta_{db}^\dagger\sigma^\mu \eta_{db}
+\eta_{ub}^\dagger\sigma^\mu \eta_{ub})\notag
\end{align}
This new link between the wave equation and the Lagrangian density is much
stronger than the old one, because it comes from a simple separation of
the different parts of a multivector in Clifford algebra. The old link, going
from the Lagrangian density to the wave equation, supposes a condition of
cancellation at infinity which is dubious in the case of a propagating wave.
On the physical point of view, there are no difficulties in the case of a
stationary wave. Difficulties begin when propagating waves are studied. Our wave 
equations, since they are compatible with an oriented time and an oriented space,
appear as more general, more physical, than Lagrangians. These are
only particular consequences of the wave equations.

On the mathematical point of view the old link is always available. It is from
the Lagrangian density (3.12) and using Lagrange equations that we have obtained 
the wave equation (1.16).

\section{Invariances}\setcounter{equation}{0}

\subsection{Form invariance of the wave equation}

Under the Lorentz dilation $R$ induced by an invertible $M$ matrix satisfying 
\begin{align}
x'&=M x M^\dagger~;~~\det(M)=r e^{i\theta}~;~~x=x^\mu\sigma_\mu~;~~x'={x'}^\mu\sigma_\mu \\
\eta_{uc} '&=\widehat{M}\eta_{uc}~;~~\eta_{dc} '=\widehat{M}\eta_{dc}~;~~\phi_{dc} ' = M\phi_{dc}
~;~~\phi_{uc} '=M \phi_{uc} \\ \Psi_c ' &=\begin{pmatrix}
\phi_{dc}' & \phi_{uc} '\\ \widehat{\phi}_{uc}' &\widehat{\phi}_{dc}'
\end{pmatrix}=\begin{pmatrix}
M & 0 \\0 & \widehat{M} \end{pmatrix}\begin{pmatrix}
\phi_{dc} & \phi_{uc} \\ \widehat{\phi}_{uc} &\widehat{\phi}_{dc}
\end{pmatrix}=N \Psi_c~;~~c=r,g,b.
\end{align}
We then let
\begin{equation}
\underline{N}=\begin{pmatrix}
N & 0 \\ 0 & N \end{pmatrix};~\underline{\pmb{\partial}}=L^\mu \partial_\mu=\begin{pmatrix}
0 & \pmb{\partial} \\ \pmb{\partial} & 0
\end{pmatrix}
\end{equation}
which implies
\begin{equation}
{{\Psi}{'}}^c=\underline{N}{\Psi}^c ;~ {{\widetilde{\Psi}}{'}}^c={\widetilde{\Psi}}^c\underline{\widetilde{N}}
;~\underline{\widetilde{N}}=\begin{pmatrix}
\widetilde N & 0 \\ 0 & \widetilde N \end{pmatrix};~\underline{\mathbf{D}}=\underline{\widetilde{N}}\ 
{\underline{\mathbf{D}}}'\underline{N}.
\end{equation}
Then we get
\begin{align}
\widetilde{\Psi}^c(\underline{\mathbf{D}}\Psi^c)L_{012}&=\widetilde{\Psi}^c\underline{\widetilde{N}}\ 
{\underline{\mathbf{D}}}'\underline{N}\Psi^c L_{012}\notag \\ &={\widetilde{\Psi}{'}}^c (
{\underline{\mathbf{D}}}'{\Psi {'}}^c ) L_{012}.
\end{align}
and we shall now study the form invariance of the mass term. All $s_j$ are determinants of a $\phi$
matrix, this implies
\begin{align}
s_j'&=\det(\phi')=\det(M\phi)=\det(M)\det(\phi)=re^{i\theta}s_j\\
{s'}_j^*&=re^{-i\theta}s_j^*;~ \rho_2 '= r \rho_2.
\end{align}
This gives
\begin{align}
{\chi{'}}^c&=\begin{pmatrix}
\chi_b ' & \chi_g ' \\ \chi_r ' & 0
\end{pmatrix}\\
r^2\rho_2^2{\chi{'}}^c&={\rho '}_2^2 {\chi{'}}^c=\begin{pmatrix}
re^{-i\theta}M & 0 \\ 0 & re^{i\theta}\widehat{M}
\end{pmatrix}\rho_2^2\chi^c \\
{\chi{'}}^c  &=\begin{pmatrix}
r^{-1}e^{-i\theta}M & 0 \\ 0 & r^{-1}e^{i\theta}\widehat{M}
\end{pmatrix}\chi_c=\widetilde{N}^{-1}\chi^c \\
{\widetilde{\Psi}{'}}^c{\chi{'}}^c&=\widetilde{\Psi}^c\widetilde{N}\widetilde{N}^{-1}\chi^c=\widetilde{\Psi}^c\chi^c
\end{align}
Then the form invariance of the wave equation is equivalent to the condition on the mass term
\begin{align}
m_2'\rho_2'&=m_2\rho_2 \\
m_2'r&=m_2
\end{align}
linked to the existence of the Planck factor \cite{dav:14}.

\subsection{Gauge invariance of the wave equation}

Since we have previously proved the gauge invariance of the lepton part of the wave equation,
it is reason enough to prove the gauge invariance of the quark part of the wave equation.

\subsubsection{Gauge group generated by $\underline{P}_0$}

We have here
\begin{align}
\underline{P}_0(\Psi^c)&=\Psi^c(-\frac{1}{3}L_{21})\\
{\Psi{'}}^c&=[\exp(\theta \underline{P}_0)](\Psi^c)=\Psi^c\exp(-\frac{\theta}{3}L_{21})\\
B_\mu '&=B_\mu-\frac{2}{g_1}B_\mu 
\end{align}
To get the gauge invariance of the wave equation we must get
\begin{equation}
{\chi{'}}^c=\chi^c\exp(-\frac{\theta}{3}L_{21});~\chi_c'=\chi_c\exp(-\frac{\theta}{3}\gamma_{21}),~c=r,g,b.
\end{equation}
This is satisfied because
\begin{align}
\phi_{dc}'&=\phi_{dc}e^{-i\frac{\theta}{3}\sigma_3};~\phi_{uc}'=\phi_{uc}e^{-i\frac{\theta}{3}\sigma_3}\\
{\eta{'}}_{1dc}^*&=e^{i\frac{\theta}{3}}\eta_{1dc}^*;~{\eta{'}}_{1uc}^*=e^{i\frac{\theta}{3}}\eta_{1uc}^*\notag\\
{\eta{'}}_{2dc}^*&=e^{i\frac{\theta}{3}}\eta_{2dc}^*;~{\eta{'}}_{2uc}^*=e^{i\frac{\theta}{3}}\eta_{2uc}^*\\
s_j'&=e^{2i\frac{\theta}{3}}s_j,~j=1,2,\dots,15.
\end{align}
All up terms in the matrix $\chi_c$ contain $s_j^*\phi_{dc}\sigma_1$ and $s_j^*\phi_{uc}\sigma_1$ terms. We get
\begin{align}
\phi_{dc}'&=\phi_{dc}e^{-i\frac{\theta}{3}\sigma_3}=e^{i\frac{\theta}{3}}\phi_{dc}\\
{s{'}}_j^*\phi_{dc}'\sigma_1 &=e^{-i\frac{\theta}{3}}\phi_{dc}\sigma_1=\phi_{dc}e^{\frac{\theta}{3}\sigma_{12}}
\sigma_1=\phi_{dc}\sigma_1e^{-\frac{\theta}{3}\sigma_{12}}\\
\chi_c'&=\chi_c\exp(-\frac{\theta}{3}\gamma_{21})\\
{\chi{'}}^c&=\chi^c\exp(-\frac{\theta}{3}L_{21}).
\end{align}
And we finally get
\begin{align}
(\underline{\mathbf{D}}'{\Psi{'}}^c)L_{012}+m_2 \rho_2'{\chi{'}}^c
&=[(\underline{\mathbf{D}}\Psi^c)L_{012}+m_2 \rho_2\chi^c]\exp(-\frac{\theta}{3}L_{21})=0
\end{align}
The wave equation with mass term is gauge invariant under the group generated by $\underline{P}_0$.

\subsubsection{Gauge group generated by $\underline{P}_1$}

We have here
\begin{align}
\underline{P}_1(\Psi^c)&=\Psi^c L_{35}\\
{\Psi{'}}^c&=[\exp(\theta \underline{P}_1)](\Psi^c)=\Psi^c\exp(\theta L_{35})\\
{W{'}}_\mu^1=W_\mu^1-\frac{2}{g_2}\partial_\mu\theta 
\end{align}
We put a more detailed calculation in C.1. We get
\begin{align}
(\underline{\mathbf{D}}'{\Psi{'}}^c)L_{012}+m_2 \rho_2'{\chi{'}}^c
&=(\underline{\mathbf{D}}\Psi^c)\exp(\theta L_{35})L_{012}+m_2 \rho_2'{\chi{'}}^c\notag \\
&=[(\underline{\mathbf{D}}\Psi^c)L_{012}+m_2 \rho_2\chi^c]\exp(\theta L_{35})=0
\end{align}
The wave equation with mass term is then gauge invariant under the group generated by $\underline{P}_1$.

\subsubsection{Gauge group generated by $\underline{P}_2$}

We have here
\begin{align}
\underline{P}_2(\Psi^c)&=\Psi^c L_{5012}\\
{\Psi{'}}^c&=[\exp(\theta \underline{P}_2)](\Psi^c)=\Psi^c\exp(\theta L_{5012})\\
{W{'}}_\mu^2=W_\mu^2-\frac{2}{g_2}\partial_\mu\theta 
\end{align}
We have put a more detailed calculation in C.2. We get
\begin{align}
(\underline{\mathbf{D}}'{\Psi{'}}^c)L_{012}+m_2 \rho_2'{\chi{'}}^c
&=(\underline{\mathbf{D}}\Psi^c)\exp(\theta L_{5012})L_{012}+m_2 \rho_2'{\chi{'}}^c\notag \\
&=[(\underline{\mathbf{D}}\Psi^c)L_{012}+m_2 \rho_2\chi^c]\exp(-\theta L_{5012})=0
\end{align}
The wave equation with mass term is then gauge invariant under the group generated by $\underline{P}_2$.

\subsubsection{Gauge group generated by $\underline{P}_3$}

We have here
\begin{align}
\underline{P}_3(\Psi^c)&=\Psi^c L_{3012}\\
{\Psi{'}}^c&=[\exp(\theta \underline{P}_3)](\Psi^c)=\Psi^c\exp(\theta L_{3012})\\
{W{'}}_\mu^3=W_\mu^3-\frac{2}{g_2}\partial_\mu\theta 
\end{align}
We have put a more detailed calculation in C.3. We get
\begin{align}
(\underline{\mathbf{D}}'{\Psi{'}}^c)L_{012}+m_2 \rho_2'{\chi{'}}^c
&=(\underline{\mathbf{D}}\Psi^c)\exp(\theta L_{3012})L_{012}+m_2 \rho_2'{\chi{'}}^c\notag \\
&=[(\underline{\mathbf{D}}\Psi^c)L_{012}+m_2 \rho_2\chi^c]\exp(-\theta L_{3012})=0
\end{align}
The wave equation with mass term is then gauge invariant under the group generated by $\underline{P}_3$.

\subsubsection{Gauge group generated by $\Gamma_1$}

We use now the gauge transformation
\begin{align}
\Psi_r'&=C \Psi_r+S\mathbf{i}\Psi_g;~C=\cos(\theta);~S=\sin(\theta)\\
\Psi_g' &=C \Psi_g+S\mathbf{i}\Psi_r\\
\Psi_b'&=\Psi_b
\end{align}
We can then forget here $\Psi_b$. The gauge invariance signifies that the system
\begin{align}
\pmb\partial \Psi_r&=-\frac{g_3}{2}\mathbf{G}^1\mathbf{i}\Psi_g+m_2\rho_2 \chi_r \gamma_{012}\notag\\
\pmb\partial \Psi_g&=-\frac{g_3}{2}\mathbf{G}^1\mathbf{i}\Psi_r+m_2\rho_2 \chi_g \gamma_{012}
\end{align}
must be equivalent to the system
\begin{align}
\pmb\partial \Psi_r'&=-\frac{g_3}{2}{\mathbf{G}{'}}^1\mathbf{i}\Psi_g'+m_2\rho_2' \chi_r' \gamma_{012}\notag\\
\pmb\partial \Psi_g'&=-\frac{g_3}{2}{\mathbf{G}{'}}^1\mathbf{i}\Psi_r'+m_2\rho_2 '\chi_g '\gamma_{012}
\end{align}
Using relations (4.39) and (4.40) the system (4.43) is equivalent to (4.42) if and only if
\begin{equation}
{\mathbf{G}{'}}^1=\mathbf{G}^1-\frac{2}{g_3}\pmb\partial\theta
\end{equation}
because we get in C.4 
\begin{align}
\rho '&=\rho \\
\chi_r'&=C\chi_r-S\mathbf{i}\chi_g \\
\chi_g'&=C\chi_g-S\mathbf{i}\chi_r
\end{align}
The change of sign of the phase between (4.39) and (4.46) comes from the anticommutation
between $\mathbf{i}$ and $\pmb\partial$.

\subsubsection{Gauge groups generated by $\Gamma_k~,~k>1$}

We use with $k=2$ the gauge transformation
\begin{align}
\Psi_r'&=C \Psi_r+S\Psi_g;~C=\cos(\theta);~S=\sin(\theta)\\
\Psi_g' &=C \Psi_g-S\Psi_r\\
\Psi_b'&=\Psi_b
\end{align}
The gauge invariance signifies that the system
\begin{align}
\pmb\partial \Psi_r&=-\frac{g_3}{2}\mathbf{G}^2\Psi_g+m_2\rho_2 \chi_r \gamma_{012}\notag\\
\pmb\partial \Psi_g&=\frac{g_3}{2}\mathbf{G}^2\Psi_r+m_2\rho_2 \chi_g \gamma_{012}
\end{align}
must be equivalent to the system
\begin{align}
\pmb\partial \Psi_r'&=-\frac{g_3}{2}{\mathbf{G}{'}}^2\Psi_g'+m_2\rho_2' \chi_r' \gamma_{012}\notag\\
\pmb\partial \Psi_g'&=\frac{g_3}{2}{\mathbf{G}{'}}^2\Psi_r'+m_2\rho_2 '\chi_g '\gamma_{012}
\end{align}
Using relations (4.48) and (4.49) the system (4.52) is equivalent to (4.51) if and only if
\begin{equation}
{\mathbf{G}{'}}^2=\mathbf{G}^2-\frac{2}{g_3}\pmb\partial\theta
\end{equation}
because we get 
\begin{align}
\rho '&=\rho \\
\chi_r'&=C\chi_r+S\chi_g \\
\chi_g'&=C\chi_g-S\chi_r.
\end{align}

The case $k=3$ is detailed in C.5 and the case $k=8$ is detailed in C.6. Cases $k=4$ and
$k=6$ are similar to $k=1$ and cases $k=5$ and $k=7$ are similar to $k=2$ by permutation
of indexes of color.

\section{Concluding remarks}

From experimental results obtained in the accelerators physicists have built what is
now known as the ``standard model". This model is generally thought to be a part of
quantum field theory, itself a part of axiomatic quantum mechanics. One of these
axioms is that each state describing a physical situation follows a Schr\"odinger
wave equation. Since this wave equation is not relativistic and does not account
for the spin 1/2 which is necessary to any fermion, the
standard model has evidently not followed the axiom and has used instead a Dirac 
equation to describe fermions. Our work
also starts with the Dirac equation. This wave equation is the linear
approximation of our nonlinear homogeneous equation of the electron.

The wave equation presented here is a wave equation for a classical wave, a function of space
and time with value into a Clifford algebra. It is not a quantized wave with value
into a Hilbertian space of operators. Nevertheless and consequently we get most of the aspects of
the standard model, for instance the fact that leptons are insensitive to
strong interactions. The standard model is much stronger than generally thought.
For instance we firstly did not use the link between the wave of the particle
and the wave of the antiparticle, but then we needed a greater Clifford algebra
and we could not get the necessary link between reversions\footnote{The reversion is
an anti-isomorphism changing the order of any product (see \cite{davi:14} 1.1). It
is specific to each Clifford algebra. The Appendix A explains the link between
the reversion in $Cl_{1,3}$ and the reversion in $Cl_{1,5}$
} that we use in our
wave equation. We also needed the existence of the inverse to build the wave
of a system of particles from the waves of its components. And we got two
general identities which exist only if all parts of the general wave are
left waves, only the electron having also a right wave. 

The most important
property of the general wave is its form invariance under a group including
the covering group of the restricted Lorentz group. Our group does not explain
why space and time are oriented, but it respects these orientations. The physical
time is then compatible with thermodynamics, and the physical space is
compatible with the violation of parity by weak interactions.
  
 The wave accounts for
all particles and anti-particles of the first generation. We have also given
\cite{davi:12}\cite{dav:12}\cite{dav:14}\cite{dabe:14} the reason of the existence
of three generations, it is simply the dimension of our physical space. Since
the $SU(3)$ gauge group of chromodynamics acts independently from the index of
generations, the physical quarks may be combinations of quarks of different generations.
Quarks composing protons and neutrons are such combinations. Our wave equation
allows only two masses at each generation, one for the lepton part of the wave,
the other one for the two quarks. The mixing can give a different mass for the two
quarks of each generation.

Since the wave equation with mass term is gauge invariant, there is no necessity
to use the mechanism of spontaneous symmetry breaking. The scalar boson certainly
exists, but it does not explain the masses.

A wave equation is only a beginning. It shall be necessary to study also the
boson part of the standard model and the systems of fermions, from this wave
equation. A construction of the wave of a system of identical particles is
possible and compatible with the Pauli principle \cite{dav2:12} \cite{davi:14}.

\begin{appendix}

\section{Calculation of the reverse in $Cl_{1,5}$}\setcounter{equation}{0}

Here indexes $\mu, \nu, \rho\dots$ have value $0,1,2,3$ and indexes $a,b,c,d,e$
have value $0,1,2,3,4,5$. We use\footnote{$I_2$, $I_4$, $I_8$ are unit matrices. The
identification process allowing to include $\mathbb{R}$ in each real Clifford algebra
allows to read $a$ instead of $aI_n$ for any complex number $a$.
} the following matrix representation of $Cl_{1,5}$:
\begin{align}
L_\mu&=\begin{pmatrix} 0 & \gamma_\mu \\ \gamma_\mu &0 \end{pmatrix} ;~
L_4=\begin{pmatrix} 0 & -I_4 \\ I_4 & 0 \end{pmatrix} ;~L_5=\begin{pmatrix}
0 &\mathbf{i} \\ \mathbf{i}&0 \end{pmatrix};~\mathbf{i}=\begin{pmatrix}iI_2 &0 
\\0 &-iI_2\end{pmatrix}\notag \\
\gamma_0 &=\gamma^0 =\begin{pmatrix} 0 & I_2 \\ I_2 & 0 \end{pmatrix} ;~\gamma_j=
-\gamma^j=\begin{pmatrix} 0 & \sigma_j \\ -\sigma_j & 0 \end{pmatrix}, ~j=1,2,3 
\end{align}
where $\sigma_j$ are Pauli matrices. This gives
\begin{align}
L_{\mu\nu}&=L_\mu L_\nu =\begin{pmatrix}
0 & \gamma_\mu \\ \gamma_\mu &0
\end{pmatrix}\begin{pmatrix}
0 & \gamma_\nu \\ \gamma_\nu &0
\end{pmatrix}=\begin{pmatrix}
\gamma_{\mu\nu}&0 \\0&\gamma_{\mu\nu}
\end{pmatrix} \\ L_{\mu\nu\rho}&=L_{\mu\nu} L_\rho =\begin{pmatrix}
\gamma_{\mu\nu} & 0 \\ 0&\gamma_{\mu\nu}\end{pmatrix}\begin{pmatrix}
0 & \gamma_\rho \\ \gamma_\rho &0
\end{pmatrix}=\begin{pmatrix}
0&\gamma_{\mu\nu\rho} \\ \gamma_{\mu\nu\rho}&0
\end{pmatrix} \\ L_{0123}&=L_{01} L_{23} =\begin{pmatrix}
\gamma_{0123} & 0 \\ 0&\gamma_{0123}\end{pmatrix}=\begin{pmatrix}
\mathbf{i}&0\\0&\mathbf{i}
\end{pmatrix}
\end{align}
We get also
\begin{align}
L_{45}&= L_4 L_5 =\begin{pmatrix}
0 & -I_4 \\I_4 &0 \end{pmatrix}\begin{pmatrix}
0 & \mathbf{i} \\\mathbf{i} &0
\end{pmatrix}=\begin{pmatrix}
-\mathbf{i}&0 \\0&\mathbf{i}
\end{pmatrix}=-L_{54} \\ L_{012345}&=\begin{pmatrix}
\mathbf{i} & 0 \\ 0&\mathbf{i}\end{pmatrix}\begin{pmatrix}
-\mathbf{i}&0 \\0&\mathbf{i}
\end{pmatrix}=\begin{pmatrix}
I_4 & 0 \\ 0&-I_4\end{pmatrix}
 \\ L_{01235}&=L_{0123}L_{5} =\begin{pmatrix}
\mathbf{i} & 0 \\ 0&\mathbf{i}\end{pmatrix}\begin{pmatrix}
0&\mathbf{i}\\\mathbf{i}&0
\end{pmatrix}=\begin{pmatrix}
0 & -I_4 \\ -I_4 &0
\end{pmatrix}.
\end{align}
Similarly we get\footnote{
$\mathbf{i}$ anti-commutes with any odd element in space-time algebra and
commutes with any even element.}
\begin{align}
L_{\mu4}&=\begin{pmatrix}
\gamma_\mu & 0 \\0 &-\gamma_\mu \end{pmatrix}
~;~~L_{\mu5}=\begin{pmatrix}
\gamma_\mu\mathbf{i} &0 \\0&\gamma_\mu\mathbf{i}
\end{pmatrix} \\ L_{\mu\nu 4}&=\begin{pmatrix}
0&-\gamma_{\mu\nu} \\ \gamma_{\mu\nu}&0 \end{pmatrix}
~;~~L_{\mu\nu 5}=\begin{pmatrix}
0&\gamma_{\mu\nu}\mathbf{i} \\ \gamma_{\mu\nu}\mathbf{i}&0 \end{pmatrix}
 \\ L_{\mu\nu\rho 4}&=\begin{pmatrix}
\gamma_{\mu\nu\rho}&0 \\0&-\gamma_{\mu\nu\rho} \end{pmatrix}
~;~~L_{\mu\nu\rho 5}=\begin{pmatrix}
\gamma_{\mu\nu\rho}\mathbf{i}&0 \\0&\gamma_{\mu\nu\rho}\mathbf{i} \end{pmatrix}
\\ L_{\mu 45}&=\begin{pmatrix}
0&\gamma_{\mu}\mathbf{i} \\-\gamma_{\mu}\mathbf{i}&0 \end{pmatrix}
~;~~L_{\mu\nu 45}=\begin{pmatrix}
-\gamma_{\mu\nu}\mathbf{i}&0 \\0&\gamma_{\mu\nu}\mathbf{i} \end{pmatrix}
\\ L_{\mu\nu\rho 45}&=\begin{pmatrix}
0&\gamma_{\mu\nu\rho}\mathbf{i} \\-\gamma_{\mu\nu\rho}\mathbf{i}&0 \end{pmatrix}
~;~~L_{01234}=\begin{pmatrix}
0&-\mathbf{i} \\\mathbf{i}&0 \end{pmatrix}
\end{align}
Scalar and pseudo-scalar terms read
\begin{align}
\alpha I_8+\omega L_{012345}&=\begin{pmatrix}
(\alpha+\omega)I_4&0 \\ 0&(\alpha-\omega)I_4
\end{pmatrix} \\ \alpha I_8-\omega \Lambda_{012345}&=\begin{pmatrix}
(\alpha-\omega)I_4&0 \\ 0&(\alpha+\omega)I_4
\end{pmatrix}
\end{align}
For the calculation of the 1-vector term $$N^a L_a=N^4 L_4+
N^5 L_5+N^\mu L_\mu$$ we let
\begin{equation}
\beta=N^4~;~~ \delta=N^5 ~;~~\mathbf{a}=N^\mu \gamma_\mu.
\end{equation}
This gives
\begin{equation}
N^a L_a=\begin{pmatrix}
0 & -\beta I_4 +\delta \mathbf{i}+\mathbf{a} \\
\beta I_4 +\delta \mathbf{i}+\mathbf{a}&0
\end{pmatrix}.
\end{equation}
For the calculation of the 2-vector term $$N^{ab} L_{ab}=N^{45} L_{45}+
N^{\mu 4} L_{\mu 4}+N^{\mu 5} L_{\mu 5}+N^{\mu\nu} L_{\mu\nu}$$ we let
\begin{equation}
\epsilon=N^{45}~;~~ \mathbf{b}=N^{\mu 4}\gamma_\mu ~;~~\mathbf{c}=N^{\mu 5}\gamma_\mu~;~~
\mathbf{A}=N^{\mu \nu}\gamma_{\mu\nu}
\end{equation}
This gives 
\begin{equation}
N^{ab} L_{ab}=\begin{pmatrix}
-\epsilon\mathbf{i}+\mathbf{b}- \mathbf{ic}+\mathbf{A} & 0\\
0 & \epsilon\mathbf{i}-\mathbf{b}- \mathbf{ic}+\mathbf{A}
\end{pmatrix}.
\end{equation}
For the calculation of the 3-vector term $$N^{abc} L_{abc}=N^{\mu 45} L_{\mu 45}+
N^{\mu\nu 4} L_{\mu\nu 4}+N^{\mu\nu 5} L_{\mu\nu 5}+N^{\mu\nu\rho} L_{\mu\nu\rho}$$ we let
\begin{equation}
\mathbf{d}=N^{\mu45}\gamma_\mu~;~~ \mathbf{B}=N^{\mu\nu 4}\gamma_{\mu\nu} ~;~~\mathbf{C}=N^{\mu\nu 5}\gamma_{\mu\nu}~;~~
\mathbf{ie}=N^{\mu \nu\rho}\gamma_{\mu\nu\rho}
\end{equation}
This gives with (A.3) and (A.9)
\begin{equation}
N^{abc}L_{abc}=\begin{pmatrix}
0&\mathbf{di}-\mathbf{B}+ \mathbf{iC}+\mathbf{ie}\\
\mathbf{id}+\mathbf{B}+ \mathbf{iC}+\mathbf{ie}&0
\end{pmatrix}.
\end{equation}
For the calculation of the 4-vector term $$N^{abcd} L_{abcd}=N^{\mu\nu 45} L_{\mu\nu 45}+
N^{\mu\nu\rho 4} L_{\mu\nu\rho 4}+N^{\mu\nu\rho 5} L_{\mu\nu\rho 5}+N^{0123} L_{0123}$$ we let
\begin{equation}
\mathbf{D}=N^{\mu\nu 45}\gamma_{\mu\nu}~;~~ \mathbf{if}=N^{\mu\nu\rho 4}\gamma_{\mu\nu\rho} 
~;~~\mathbf{ig}=N^{\mu\nu\rho 5}\gamma_{\mu\nu\rho}~;~~\zeta=N^{0123}
\end{equation}
This gives with (A.4) and (A.10)
\begin{equation}
N^{abcd}L_{abcd}=\begin{pmatrix}
-\mathbf{iD}+\mathbf{if}+ \mathbf{g}+\zeta\mathbf{i}&0\\
0& \mathbf{iD}-\mathbf{if}+ \mathbf{g}+\zeta\mathbf{i}
\end{pmatrix}.
\end{equation}
For the calculation of the pseudo-vector term $$N^{abcde} L_{abcde}=N^{\mu\nu\rho 45} L_{\mu\nu\rho 45}+
N^{01234} L_{01234}+N^{01235} L_{01235}$$ we let
\begin{equation}
\mathbf{ih}=N^{\mu\nu\rho 45}\gamma_{\mu\nu\rho}~;~~ \eta=N^{01234} 
~;~~\theta=N^{01235}
\end{equation}
This gives with (A.7) and (A.12)
\begin{equation}
N^{abcde}L_{abcde}=\begin{pmatrix}
0& \mathbf{h}-\eta\mathbf{i}- \theta I_4\\
 -\mathbf{h}+\eta\mathbf{i}- \theta I_4
\end{pmatrix}.
\end{equation}
We then get
\begin{align}
&\Psi= \begin{pmatrix}
\Psi_l & \Psi_r \\ \Psi_g &\Psi_b
\end{pmatrix}\\
&=\begin{pmatrix}
(\alpha+\omega)I_4 +( \mathbf{b}+\mathbf{g})+(\mathbf{A}-\mathbf{iD}) &
-(\beta+\theta)I_4+(\mathbf{a}+\mathbf{h})+(-\mathbf{B}+\mathbf{iC}) \\ 
+\mathbf{i}(-\mathbf{c}+\mathbf{f}) +(\zeta-\epsilon)\mathbf{i} & +\mathbf{i}(-\mathbf{d}+\mathbf{e})
+(\delta-\eta)\mathbf{i} \\
\\
(\beta-\theta)I_4+(\mathbf{a}-\mathbf{h})+(\mathbf{B}+\mathbf{iC}) &
(\alpha-\omega)I_4 +(-\mathbf{b}+\mathbf{g})+(\mathbf{A}+\mathbf{iD}) \\ +\mathbf{i}(\mathbf{d}+\mathbf{e})
+(\delta+\eta)\mathbf{i} &+\mathbf{i}(-\mathbf{c}-\mathbf{f})
+(\zeta+\epsilon)\mathbf{i} 
\end{pmatrix}\notag
\end{align}
This implies
\begin{align}
\Psi_l &= (\alpha+\omega) +(\mathbf{b}+\mathbf{g})+(\mathbf{A}-\mathbf{iD}) 
+\mathbf{i}(-\mathbf{c}+\mathbf{f}) +(\zeta-\epsilon)\mathbf{i} \\ 
\Psi_r &=-(\beta+\theta)+(\mathbf{a}+\mathbf{h})+(-\mathbf{B}+\mathbf{iC})
+\mathbf{i}(-\mathbf{d}+\mathbf{e}) +(\delta-\eta)\mathbf{i} \\
\Psi_g &=(\beta-\theta)+(\mathbf{a}-\mathbf{h})+(\mathbf{B}+\mathbf{iC})
+\mathbf{i}(\mathbf{d}+\mathbf{e}) +(\delta+\eta)\mathbf{i} \\
\Psi_b &= (\alpha-\omega) +(-\mathbf{b}+\mathbf{g})+(\mathbf{A}+\mathbf{iD}) 
+\mathbf{i}(-\mathbf{c}-\mathbf{f}) +(\zeta+\epsilon)\mathbf{i}  
\end{align}
In $Cl_{1,3}$ the reverse of $$A=<A>_0+<A>_1+<A>_2+<A>_3+<A>_4$$ is
$$\widetilde A=<A>_0+<A>_1-<A>_2-<A>_3+<A>_4$$ we must change the sign of bivectors 
$\mathbf{A}$, $\mathbf{B}$,
$\mathbf{iC}$, $\mathbf{iD}$, and trivectors $\mathbf{ic}$, $\mathbf{id}$, $\mathbf{ie}$,
$\mathbf{if}$ and we then get
\begin{align}
\widetilde \Psi_l &= (\alpha+\omega) +(\mathbf{b}+\mathbf{g})+(-\mathbf{A}+\mathbf{iD}) 
+\mathbf{i}(\mathbf{c}-\mathbf{f}) +(\zeta-\epsilon)\mathbf{i} \\ 
\widetilde \Psi_r &=-(\beta+\theta)+(\mathbf{a}+\mathbf{h})+(\mathbf{B}-\mathbf{iC})
+\mathbf{i}(\mathbf{d}-\mathbf{e}) +(\delta-\eta)\mathbf{i} \\
\widetilde \Psi_g &=(\beta-\theta)+(\mathbf{a}-\mathbf{h})-(\mathbf{B}+\mathbf{iC})
-\mathbf{i}(\mathbf{d}+\mathbf{e}) +(\delta+\eta)\mathbf{i} \\
\widetilde \Psi_b &= (\alpha-\omega)I_4 +(- \mathbf{b}+\mathbf{g})-(\mathbf{A}+\mathbf{iD}) 
+\mathbf{i}(\mathbf{c}+\mathbf{f}) +(\zeta+\epsilon)\mathbf{i}  
\end{align}
The reverse, in $Cl_{1,5}$ now, of $$A=<A>_0+<A>_1+<A>_2+<A>_3+<A>_4+<A>_5+<A>_6$$ is
$$\widetilde A=<A>_0+<A>_1-<A>_2-<A>_3+<A>_4+<A>_5-<A>_6$$
Only terms which change sign, with (A.13),  (A.18) and (A.20), are scalars $\epsilon$ and 
$\omega$, vectors $\mathbf{b}$,
$\mathbf{c}$, $\mathbf{d}$, $\mathbf{e}$ and bivectors $\mathbf{A}$,
$\mathbf{B}$, $\mathbf{C}$. These changes of sign are not the same in $Cl_{1,5}$
as in $Cl_{1,3}$. Differences are corrected by the fact that the reversion in $Cl_{1,5}$
also exchanges the place of $\Psi_l$ and $\Psi_b$ terms. We then get from (A.25)
\begin{align}
\widetilde\Psi&= 
\begin{pmatrix}
(\alpha-\omega)I_4 +(-\mathbf{b}+\mathbf{g})+(-\mathbf{A}-\mathbf{iD}) &
-(\beta+\theta)I_4+(\mathbf{a}+\mathbf{h})+(\mathbf{B}-\mathbf{iC}) \\ 
+\mathbf{i}(\mathbf{c}+\mathbf{f}) +(\zeta+\epsilon)\mathbf{i} & +\mathbf{i}(\mathbf{d}-\mathbf{e})
+(\delta-\eta)\mathbf{i} \\
\\
(\beta-\theta)I_4+(\mathbf{a}-\mathbf{h})-(\mathbf{B}+\mathbf{iC}) &
(\alpha+\omega)I_4 +(\mathbf{b}+\mathbf{g})+(-\mathbf{A}+\mathbf{iD}) \\ -\mathbf{i}(\mathbf{d}+\mathbf{e})
+(\delta+\eta)\mathbf{i} &+\mathbf{i}(\mathbf{c}-\mathbf{f})
+(\zeta-\epsilon)\mathbf{i} 
\end{pmatrix}\notag\\
&=\begin{pmatrix}
\widetilde\Psi_b & \widetilde\Psi_r \\ \widetilde\Psi_g &\widetilde\Psi_l
\end{pmatrix}.
\end{align}
This link between the reversion in $Cl_{1,3}$ and the reversion in $Cl_{1,5}$ is necessary to get an
invariant wave equation. It is not general, for instance the reversion in $Cl_3$ is
not linked to the reversion in $Cl_{2,3}$.

\section{Scalar densities and $\chi$ terms}\setcounter{equation}{0}

There are $6\times 5/2=15$ such complex scalar densities:
\begin{align}
s_1&=2(\xi_{1\overline{u}r}\eta_{1ug}^*+\xi_{2\overline{u}r}\eta_{2ug}^*)=
2(\eta_{2ur}^*\eta_{1ug}^*-\eta_{1ur}^*\eta_{2ug}^*)\\
s_2&=2(\xi_{1\overline{u}g}\eta_{1ub}^*+\xi_{2\overline{u}g}\eta_{2ub}^*)=
2(\eta_{2ug}^*\eta_{1ub}^*-\eta_{1ug}^*\eta_{2ub}^*)\\
s_3&=-2(\xi_{1\overline{u}r}\eta_{1ub}^*+\xi_{2\overline{u}r}\eta_{2ub}^*)=
2(\eta_{2ub}^*\eta_{1ur}^*-\eta_{1ub}^*\eta_{2ur}^*)
\end{align}
\begin{align}
s_4&=2(\xi_{1\overline{d}r}\eta_{1dg}^*+\xi_{2\overline{d}r}\eta_{2dg}^*)=
2(\eta_{2dr}^*\eta_{1dg}^*-\eta_{1dr}^*\eta_{2dg}^*)\\
s_5&=2(\xi_{1\overline{d}g}\eta_{1db}^*+\xi_{2\overline{d}g}\eta_{2db}^*)=
2(\eta_{2dg}^*\eta_{1db}^*-\eta_{1dg}^*\eta_{2db}^*)\\
s_6&=-2(\xi_{1\overline{d}r}\eta_{1db}^*+\xi_{2\overline{d}r}\eta_{2db}^*)=
2(\eta_{2db}^*\eta_{1dr}^*-\eta_{1db}^*\eta_{2dr}^*)
\end{align}
\begin{align}
s_7&=2(\xi_{1\overline{u}r}\eta_{1dr}^*+\xi_{2\overline{u}r}\eta_{2dr}^*)=
2(\eta_{2ur}^*\eta_{1dr}^*-\eta_{1ur}^*\eta_{2dr}^*)\\
s_8&=2(\xi_{1\overline{u}g}\eta_{1dg}^*+\xi_{2\overline{u}g}\eta_{2dg}^*)=
2(\eta_{2ug}^*\eta_{1dg}^*-\eta_{1ug}^*\eta_{2dg}^*)\\
s_9&=2(\xi_{1\overline{u}b}\eta_{1db}^*+\xi_{2\overline{u}b}\eta_{2db}^*)=
2(\eta_{2ub}^*\eta_{1db}^*-\eta_{1ub}^*\eta_{2db}^*)
\end{align}
\begin{align}
s_{10}&=2(\xi_{1\overline{u}r}\eta_{1dg}^*+\xi_{2\overline{u}r}\eta_{2dg}^*)=
2(\eta_{2ur}^*\eta_{1dg}^*-\eta_{1ur}^*\eta_{2dg}^*)\\
s_{11}&=2(\xi_{1\overline{u}g}\eta_{1db}^*+\xi_{2\overline{u}g}\eta_{2db}^*)=
2(\eta_{2ug}^*\eta_{1db}^*-\eta_{1ug}^*\eta_{2db}^*)\\
s_{12}&=-2(\xi_{1\overline{d}r}\eta_{1ub}^*+\xi_{2\overline{d}r}\eta_{2ub}^*)=
2(\eta_{2ub}^*\eta_{1dr}^*-\eta_{1ub}^*\eta_{2dr}^*)
\end{align}
\begin{align}
s_{13}&=2(\xi_{1\overline{u}r}\eta_{1db}^*+\xi_{2\overline{u}r}\eta_{2db}^*)=
2(\eta_{2ur}^*\eta_{1db}^*-\eta_{1ur}^*\eta_{2db}^*)\\
s_{14}&=-2(\xi_{1\overline{d}r}\eta_{1ug}^*+\xi_{2\overline{d}r}\eta_{2ug}^*)=
2(\eta_{2ug}^*\eta_{1dr}^*-\eta_{1ug}^*\eta_{2dr}^*)\\
s_{15}&=-2(\xi_{1\overline{d}g}\eta_{1ub}^*+\xi_{2\overline{d}g}\eta_{2ub}^*)=
2(\eta_{2ub}^*\eta_{1dg}^*-\eta_{1ub}^*\eta_{2dg}^*).
\end{align}
We used in \cite{dabe:14}
\begin{equation}
\chi_l=\frac{1}{\rho_1^2}\begin{pmatrix}
a_1^* \phi_e +a_2^*\phi_n\sigma_1+a_3^* \phi_n & -a_2^*\phi_{eL}\sigma_1+a_3^*\phi_{eR} \\
a_2\widehat \phi_{eL}\sigma_1 + a_3\widehat \phi_{eR} & a_1\widehat \phi_e-a_2\widehat\phi_n
\sigma_1 +a_3\widehat \phi_n
\end{pmatrix}
\end{equation}
with $\phi_{eR}=\phi_e(1+\sigma_3)/2$ and $\phi_{eL}=\phi_e(1-\sigma_3)/2$, and we need now
\begin{align}
\rho_2^2 \chi_r=
&\begin{pmatrix}
\begin{pmatrix}
s_4^*\phi_{dg}-s_6^*\phi_{db}-s_7^*\phi_{ur}\\-s_{12}^*\phi_{ub}-s_{14}^*\phi_{ug}
\end{pmatrix} \sigma_1 &\begin{pmatrix}s_1^*\phi_{ug}-s_3^*\phi_{ub}+s_7^*\phi_{dr}\\
+s_{10}^*\phi_{dg}+s_{13}^*\phi_{db} \end{pmatrix}
\sigma_1\\ \begin{pmatrix}-s_1\widehat{\phi}_{ug}+s_3\widehat{\phi}_{ub}-s_7\widehat{\phi}_{dr}\\
-s_{10}\widehat{\phi}_{dg}-s_{13}\widehat{\phi}_{db} \end{pmatrix}\sigma_1&
\begin{pmatrix}-s_4\widehat{\phi}_{dg}+s_6\widehat{\phi}_{db}+s_7\widehat{\phi}_{ur} \\ 
+s_{12}\widehat{\phi}_{ub}+s_{14}\widehat{\phi}_{ug}\end{pmatrix}\sigma_1
\end{pmatrix}\\
\rho_2^2 \chi_g=
&\begin{pmatrix}
\begin{pmatrix}
s_5^*\phi_{db}-s_4^*\phi_{dr}-s_8^*\phi_{ug}\\-s_{10}^*\phi_{ur}-s_{15}^*\phi_{ub}
\end{pmatrix} \sigma_1 &\begin{pmatrix}s_2^*\phi_{ub}-s_1^*\phi_{ur}+s_8^*\phi_{dg}\\
+s_{11}^*\phi_{db}+s_{14}^*\phi_{dr} \end{pmatrix}
\sigma_1\\ \begin{pmatrix}-s_2\widehat{\phi}_{ub}+s_1\widehat{\phi}_{ur}-s_8\widehat{\phi}_{dg}\\
-s_{11}\widehat{\phi}_{db}-s_{14}\widehat{\phi}_{dr} \end{pmatrix}\sigma_1&
\begin{pmatrix}-s_5\widehat{\phi}_{db}+s_4\widehat{\phi}_{dr}+s_8\widehat{\phi}_{ug} \\ 
+s_{10}\widehat{\phi}_{ur}+s_{15}\widehat{\phi}_{ub}\end{pmatrix}\sigma_1
\end{pmatrix}\\
\rho_2^2 \chi_b=
&\begin{pmatrix}
\begin{pmatrix}
s_6^*\phi_{dr}-s_5^*\phi_{dg}-s_9^*\phi_{ub}\\-s_{11}^*\phi_{ug}-s_{13}^*\phi_{ur}
\end{pmatrix} \sigma_1 &\begin{pmatrix}s_3^*\phi_{ur}-s_2^*\phi_{ug}+s_9^*\phi_{db}\\
+s_{12}^*\phi_{dr}+s_{15}^*\phi_{dg} \end{pmatrix}
\sigma_1\\ \begin{pmatrix}-s_3\widehat{\phi}_{ur}+s_2\widehat{\phi}_{ug}-s_9\widehat{\phi}_{db}\\
-s_{12}\widehat{\phi}_{dr}-s_{15}\widehat{\phi}_{dg} \end{pmatrix}\sigma_1&
\begin{pmatrix}-s_6\widehat{\phi}_{dr}+s_5\widehat{\phi}_{dg}+s_9\widehat{\phi}_{ub} \\ 
+s_{11}\widehat{\phi}_{ug}+s_{13}\widehat{\phi}_{ur}\end{pmatrix}\sigma_1
\end{pmatrix}\end{align}

\section{Gauge invariance, details}\setcounter{equation}{0}

\subsection{Gauge group generated by $\underline{P}_1$}

Since $\underline{P}_1(\Psi^c)=\Psi^c L_{35}$ we get
\begin{align}
{\Psi{'}}^c&=[\exp(\theta\underline{P}_1)](\Psi^c)=\Psi^c\exp(\theta L_{35}) \\
\Psi_c'&=\Psi_c e^{\theta \gamma_3\mathbf{i}},~c=r,g,b.
\end{align}
We let
\begin{equation}
C=\cos(\theta)~;~~S=\sin(\theta)
\end{equation}
Then (C.2) is equivalent to the system
\begin{align}
\widehat{\phi}_{dc}'&=C\widehat{\phi}_{dc}-iS\widehat{\phi}_{uc}\sigma_3 \\
\widehat{\phi}_{uc}'&=C\widehat{\phi}_{uc}-iS\widehat{\phi}_{dc}\sigma_3 
\end{align}
or to the system
\begin{align}
\eta_{1dc}'&=C\eta_{1dc}-iS\eta_{1uc};~{\eta{'}}_{1dc}^*=C\eta_{1dc}^*+iS\eta_{1uc}^* \\
\eta_{2dc}'&=C\eta_{2dc}-iS\eta_{2uc};~{\eta{'}}_{2dc}^*=C\eta_{2dc}^*+iS\eta_{2uc}^* \\
\eta_{1uc}'&=C\eta_{1uc}-iS\eta_{1dc};~{\eta{'}}_{1uc}^*=C\eta_{1uc}^*+iS\eta_{1dc}^* \\
\eta_{2uc}'&=C\eta_{2uc}-iS\eta_{2dc};~{\eta{'}}_{2uc}^*=C\eta_{2uc}^*+iS\eta_{2dc}^* 
\end{align}
We then get
\begin{align}
s_1'&=C^2s_1-S^2s_4+iCS(s_{10}-s_{14})\\
s_4'&=C^2s_4-S^2s_1+iCS(s_{10}-s_{14})\\
s_{10}'&=C^2s_{10}+S^2s_{14}+iCS(s_1+s_4) \\
s_{14}'&=C^2s_{14}+S^2s_{10}-iCS(s_1+s_4).
\end{align}
This implies
\begin{equation}
s_{1}'{s{'}}_{1}^* + s_{4}'{s{'}}_{4}^* + s_{10}'{s{'}}_{10}^*+ s_{14}'{s{'}}_{14}^*
=s_1 s_1^* + s_4 s_4^* + s_{10}s_{10}^*+ s_{14}s_{14}^*.
\end{equation}
Similarly, permuting colors, we get
\begin{align}
s_2'&=C^2s_2-S^2s_5+iCS(s_{11}-s_{15})\\
s_5'&=C^2s_5-S^2s_2+iCS(s_{11}-s_{15})\\
s_{11}'&=C^2s_{11}+S^2s_{15}+iCS(s_2+s_5) \\
s_{15}'&=C^2s_{15}+S^2s_{11}-iCS(s_2+s_5).
\end{align}
This implies
\begin{equation}
s_{2}'{s{'}}_{2}^* + s_{5}'{s{'}}_{5}^* + s_{11}'{s{'}}_{11}^*+ s_{15}'{s{'}}_{15}^*
=s_2 s_2^* + s_5 s_5^* + s_{11}s_{11}^*+ s_{15}s_{15}^*.
\end{equation}
and also
\begin{align}
s_3'&=C^2s_3-S^2s_6+iCS(s_{12}-s_{13})\\
s_6'&=C^2s_6-S^2s_3+iCS(s_{12}-s_{13})\\
s_{12}'&=C^2s_{12}+S^2s_{13}+iCS(s_3+s_6) \\
s_{13}'&=C^2s_{13}+S^2s_{12}-iCS(s_3+s_6).
\end{align}
This implies
\begin{equation}
s_{3}'{s{'}}_{3}^* + s_{6}'{s{'}}_{6}^* + s_{12}'{s{'}}_{12}^*+ s_{13}'{s{'}}_{13}^*
=s_3 s_3^* + s_6 s_6^* + s_{12}s_{12}^*+ s_{13}s_{13}^*.
\end{equation}
Moreover we get
\begin{equation}
s_7'=s_7;~s_8'=s_8;~s_9'=s_9.
\end{equation}
We then get
\begin{equation}
\rho'=\rho
\end{equation}
Next we have
\begin{align}
\chi_r&=\begin{pmatrix}
A & B \\\widehat{B}&\widehat{A}\end{pmatrix};~\chi_r'=\begin{pmatrix}
A' & B' \\\widehat{B}'&\widehat{A}'\end{pmatrix}\\
\widehat{A}&=(-s_4 \widehat{\phi}_{dg}+s_6 \widehat{\phi}_{db}+s_7 \widehat{\phi}_{ur}
+s_{12} \widehat{\phi}_{ub}+s_{14} \widehat{\phi}_{ug})\sigma_1 \\
\widehat{B}&=(-s_1 \widehat{\phi}_{ug}+s_3 \widehat{\phi}_{ub}-s_7 \widehat{\phi}_{dr}
-s_{10} \widehat{\phi}_{dg}-s_{13} \widehat{\phi}_{db})\sigma_1.
\end{align}
and we get
\begin{align}
\widehat{A}'&=C\widehat{A}-iS\widehat{B}\sigma_3 \\
\widehat{B}'&=C\widehat{B}-iS\widehat{A}\sigma_3 \\
\chi_r'&=\chi_r\begin{pmatrix}
C & -iS\sigma_3 \\-iS\sigma_3 & C
\end{pmatrix}=\chi_re^{\theta\gamma_3\mathbf{i}}.
\end{align}
Since we get the same relation for g and b colors we finally get
\begin{equation}
{\chi{'}}^c=\chi^c \exp(\theta L_{35})
\end{equation}

\subsection{Gauge group generated by $\underline{P}_2$}

Since $\underline{P}_2(\Psi^c)=\Psi^c L_{5012}$ we get
\begin{align}
{\Psi{'}}^c&=[\exp(\theta\underline{P}_2)](\Psi^c)=\Psi^c\exp(\theta L_{5012}) \\
\Psi_c'&=\Psi_c e^{\theta \gamma_3},~c=r,g,b.
\end{align}
We let
\begin{equation}
C=\cos(\theta)~;~~S=\sin(\theta)
\end{equation}
Then (C.35) is equivalent to the system
\begin{align}
\widehat{\phi}_{dc}'&=C\widehat{\phi}_{dc}+S\widehat{\phi}_{uc} \\
\widehat{\phi}_{uc}'&=C\widehat{\phi}_{uc}-S\widehat{\phi}_{dc} 
\end{align}
or to the system
\begin{align}
\eta_{1dc}'&=C\eta_{1dc}+S\eta_{1uc};~{\eta{'}}_{1dc}^*=C\eta_{1dc}^*+S\eta_{1uc}^* \\
\eta_{2dc}'&=C\eta_{2dc}+S\eta_{2uc};~{\eta{'}}_{2dc}^*=C\eta_{2dc}^*+S\eta_{2uc}^* \\
\eta_{1uc}'&=C\eta_{1uc}-S\eta_{1dc};~{\eta{'}}_{1uc}^*=C\eta_{1uc}^*-S\eta_{1dc}^* \\
\eta_{2uc}'&=C\eta_{2uc}-S\eta_{2dc};~{\eta{'}}_{2uc}^*=C\eta_{2uc}^*-S\eta_{2dc}^* 
\end{align}
We then get
\begin{align}
s_1'&=C^2s_1+S^2s_4-CS s_{10}+CS s_{14}\\
s_4'&=C^2s_4+S^2s_1+CS s_{10}-CS s_{14}\\
s_{10}'&=C^2 s_{10}+S^2 s_{14}+CS s_1-CS s_4 \\
s_{14}'&=C^2 s_{14}+S^2 s_{10}-CS s_1+CS s_4.
\end{align}
This implies
\begin{equation}
s_{1}'{s{'}}_{1}^* + s_{4}'{s{'}}_{4}^* + s_{10}'{s{'}}_{10}^*+ s_{14}'{s{'}}_{14}^*
=s_1 s_1^* + s_4 s_4^* + s_{10}s_{10}^*+ s_{14}s_{14}^*.
\end{equation}
Similarly, permuting colors, we get
\begin{align}
s_2'&=C^2s_2+S^2s_5-CS s_{11}+CS s_{15}\\
s_5'&=C^2s_5+S^2s_2+CS s_{11}-CS s_{15}\\
s_{11}'&=C^2 s_{11}+S^2 s_{15}+CS s_2-CS s_5 \\
s_{15}'&=C^2 s_{15}+S^2 s_{11}-CS s_2+CS s_5.
\end{align}
This implies
\begin{equation}
s_{2}'{s{'}}_{2}^* + s_{5}'{s{'}}_{5}^* + s_{11}'{s{'}}_{11}^*+ s_{15}'{s{'}}_{15}^*
=s_2 s_2^* + s_5 s_5^* + s_{11}s_{11}^*+ s_{15}s_{15}^*.
\end{equation}
and also
\begin{align}
s_3'&=C^2s_3+S^2s_6-CS s_{12}+CS s_{13}\\
s_6'&=C^2s_6+S^2s_3+CS s_{12}-CS s_{13}\\
s_{12}'&=C^2 s_{12}+S^2 s_{13}+CS s_3-CS s_6 \\
s_{13}'&=C^2 s_{13}+S^2 s_{12}-CS s_3+CS s_6.
\end{align}
This implies
\begin{equation}
s_{3}'{s{'}}_{3}^* + s_{6}'{s{'}}_{6}^* + s_{12}'{s{'}}_{12}^*+ s_{13}'{s{'}}_{13}^*
=s_3 s_3^* + s_6 s_6^* + s_{12}s_{12}^*+ s_{13}s_{13}^*.
\end{equation}
Moreover we get
\begin{equation}
s_7'=s_7;~s_8'=s_8;~s_9'=s_9.
\end{equation}
We then get
\begin{equation}
\rho'=\rho
\end{equation}
Next we get with (C.27)
\begin{align}
\widehat{A}'&=C\widehat{A}-S\widehat{B}\sigma_3 \\
\widehat{B}'&=C\widehat{B}+S\widehat{A}\sigma_3 \\
\chi_r'&=\chi_r\begin{pmatrix}
C & -S\sigma_3 \\ S\sigma_3 & C
\end{pmatrix}=\chi_re^{-\theta\gamma_3}.
\end{align}
Since we get the same relation for g and b colors we finally get
\begin{equation}
{\chi{'}}^c=\chi^c \exp(-\theta L_{5012})
\end{equation}
\subsection{Gauge group generated by $\underline{P}_3$}

Since $\underline{P}_3(\Psi^c)=\Psi^c L_{3012}$ we get
\begin{align}
{\Psi{'}}^c&=[\exp(\theta\underline{P}_3)](\Psi^c)=\Psi^c\exp(\theta L_{3012}) \\
\Psi_c'&=\Psi_c e^{\theta \gamma_{3012}},~c=r,g,b.
\end{align}
Then (C.65) is equivalent to the system
\begin{align}
\widehat{\phi}_{dc}'&=e^{i\theta}\widehat{\phi}_{dc} \\
\widehat{\phi}_{uc}'&=e^{-i\theta}\widehat{\phi}_{uc}
\end{align}
or to the system
\begin{align}
\eta_{1dc}'&=e^{i\theta}\eta_{1dc};~{\eta{'}}_{1dc}^*=e^{-i\theta}\eta_{1dc}^* \\
\eta_{2dc}'&=e^{i\theta}\eta_{2dc};~{\eta{'}}_{2dc}^*=e^{-i\theta}\eta_{2dc}^* \\
\eta_{1uc}'&=e^{-i\theta}\eta_{1uc};~{\eta{'}}_{1uc}^*=e^{i\theta}\eta_{1uc}^* \\
\eta_{2uc}'&=e^{-i\theta}\eta_{2uc};~{\eta{'}}_{2uc}^*=e^{i\theta}\eta_{2uc}^* 
\end{align}
We then get
\begin{align}
s_1'&=e^{2i\theta}s_1;~s_2'=e^{2i\theta}s_2;~s_3'=e^{2i\theta}s_3 \\
s_4'&=e^{-2i\theta}s_4;~s_5'=e^{-2i\theta}s_5;~s_6'=e^{-2i\theta}s_6 \\
s_{7}'&=s_{7};~s_{8}'=s_{8};~s_{9}'=s_{9} \\
s_{10}'&=s_{10};~s_{11}'=s_{11};~s_{12}'=s_{12} \\
s_{13}'&=s_{13};~s_{14}'=s_{14};~s_{15}'=s_{15} .
\end{align}
This implies
\begin{equation}
\rho'=\rho
\end{equation}
Next we get with (C.27)
\begin{align}
\widehat{A}'&=e^{-i\theta}\widehat{A}~;~~A'=e^{i\theta}A \\
\widehat{B}'&=e^{i\theta}\widehat{B}~;~~B'=e^{-i\theta}B \\
\chi_r'&=\chi_r\begin{pmatrix}
e^{i\theta} & 0 \\ 0 & e^{-i\theta}
\end{pmatrix}=\chi_re^{\theta\mathbf{i}}.
\end{align}
Since we get the same relation for g and b colors we finally get
\begin{equation}
{\chi{'}}^c=\chi^c \exp(-\theta L_{3012})
\end{equation}

\subsection{Gauge group generated by $\underline{\mathbf{i}}\Gamma_1$}

We name $f_1$ the gauge transformation
\begin{equation}
f_1:\Psi^c \mapsto \underline{\mathbf{i}}\Gamma_1(\Psi^c)=\begin{pmatrix}
0 & \mathbf{i} \Psi_g \\\mathbf{i} \Psi_r & 0
\end{pmatrix}
\end{equation}
which implies with $C=\cos(\theta)$ and $S=\sin(\theta)$
\begin{align}
[\exp(\theta f_1)](\Psi^c)&=\begin{pmatrix}
0 & C\Psi_r +S\mathbf{i}\Psi_g \\ C\Psi_g +S\mathbf{i}\Psi_r & \Psi_b
\end{pmatrix}=\begin{pmatrix}
0 & \Psi_r' \\ \Psi_g ' & \Psi_b '
\end{pmatrix}\\
\Psi_r'&=C\Psi_r +S\mathbf{i}\Psi_g \\
\Psi_g '&=C\Psi_g +S\mathbf{i}\Psi_r \\
\Psi_b '&= \Psi_b
\end{align}
The equality (C.84) is equivalent to the system
\begin{align}
{\eta{'}}_{1dr}^*&=C\eta_{1dr}^*+iS\eta_{1dg}^*;~{\eta{'}}_{1ur}^*=C\eta_{1ur}^*+iS\eta_{1ug}^*\\
{\eta{'}}_{2dr}^*&=C\eta_{2dr}^*+iS\eta_{2dg}^*;~{\eta{'}}_{2ur}^*=C\eta_{2ur}^*+iS\eta_{2ug}^*
\end{align}
The equality (C.85) is equivalent to the system
\begin{align}
{\eta{'}}_{1dg}^*&=C\eta_{1dg}^*+iS\eta_{1dr}^*;~{\eta{'}}_{1ug}^*=C\eta_{1ug}^*+iS\eta_{1ur}^*\\
{\eta{'}}_{2dg}^*&=C\eta_{2dg}^*+iS\eta_{2dr}^*;~{\eta{'}}_{2ug}^*=C\eta_{2ug}^*+iS\eta_{2ur}^*
\end{align}
This gives for the invariant scalars $s_j$
\begin{align}
s_1'&=s_1;~s_4'=s_4;~s_9'=s_9 \\
s_2'&=Cs_2-iSs_3;~s_3'=Cs_3-iSs_2 \\
s_5'&=Cs_5-iSs_6;~s_6'=Cs_6-iSs_5 \\
s_{11}'&=Cs_{11}+iSs_{13};~s_{13}'=Cs_{13}+iSs_{11} \\
s_{12}'&=Cs_{12}+iSs_{15};~s_{15}'=Cs_{15}+iSs_{12}
\end{align}
\begin{align}
s_7'&=C^2 s_7 - S^2 s_8 + iCS s_{10}+iCS s_{14} \\
s_8'&=C^2 s_8 - S^2 s_7 + iCS s_{14}+iCS s_{10}\\
s_{10}'&=C^2 s_{10}-S^2 s_{14}+iCS s_7 +iCS s_8 \\
s_{14}'&=C^2 s_{14}-S^2 s_{10}+iCS s_8 +iCS s_7   
\end{align}
We then get
\begin{align}
s_2'{s{'}}_2^*+s_3'{s{'}}_3^*&=s_2 s_2^* + s_3 s_3^* \\
s_5'{s{'}}_5^*+s_6'{s{'}}_6^*&=s_5 s_5^* + s_6 s_6^*\\
s_{11}'{s{'}}_{11}^*+s_{13}'{s{'}}_{13}^*&=s_{11} s_{11}^* + s_{13} s_{13}^* \\
s_{12}'{s{'}}_{12}^*+s_{15}'{s{'}}_{15}^*&=s_{12} s_{12}^* + s_{15} s_{15}^* \\
s_7'{s{'}}_7^*+s_8'{s{'}}_8^*+s_{10}'{s{'}}_{10}^*+s_{14}'{s{'}}_{14}^*&=
s_7 s_7^* + s_8 s_8^*+s_{10} s_{10}^* + s_{14} s_{14}^*\\
\rho'&=\rho.
\end{align}
Next we let
\begin{align}
\chi_r&=\begin{pmatrix} A_r & B_r \\ \widehat{B}_r & \widehat{A}_r \end{pmatrix} ;~
\chi_r'=\begin{pmatrix} A_r' & B_r' \\ \widehat{B}_r' & \widehat{A}_r' \end{pmatrix} \\
\chi_g&=\begin{pmatrix} A_g & B_g \\ \widehat{B}_g & \widehat{A}_g \end{pmatrix} ;~
\chi_g'=\begin{pmatrix} A_g' & B_g' \\ \widehat{B}_g' & \widehat{A}_g' \end{pmatrix} 
\end{align}
and we get with (B.17) and (B.18)
\begin{align}
A_r'&=C A_r -i S A_g;~B_r'=C B_r -i S B_g \\
A_g'&= CA_g-iS A_r;~B_g'=C B_g -i S B_r.
\end{align}
This gives the awaited result
\begin{equation}
\chi_r'=C\chi_r-\mathbf{i} S \chi_g;~\chi_g'=C\chi_g-\mathbf{i}S\chi_r.
\end{equation}

\subsection{Gauge group generated by $\underline{\mathbf{i}}\Gamma_3$}

We name $f_3$ the gauge transformation
\begin{equation}
f_3:\Psi^c \mapsto \underline{\mathbf{i}}\Gamma_3(\Psi^c)=\begin{pmatrix}
0 & \mathbf{i} \Psi_r \\-\mathbf{i} \Psi_g & 0
\end{pmatrix}
\end{equation}
which implies
\begin{align}
[\exp(\theta f_1)](\Psi^c)&=\begin{pmatrix}
0 & e^{\theta\mathbf{i}}\Psi_r  \\ e^{-\theta\mathbf{i}}\Psi_g  & \Psi_b
\end{pmatrix}=\begin{pmatrix}
0 & \Psi_r' \\ \Psi_g ' & \Psi_b '
\end{pmatrix}\\
\Psi_r'&=e^{\theta\mathbf{i}}\Psi_r \\
\Psi_g '&=e^{-\theta\mathbf{i}}\Psi_g \\
\Psi_b '&= \Psi_b
\end{align}
The equality (C.113) is equivalent to
\begin{align}
\begin{pmatrix}
\phi_{dr}' & \phi_{ur}' \\ \widehat{\phi}_{ur}'& \widehat{\phi}_{dr}'
\end{pmatrix}&=\begin{pmatrix} e^{i\theta}&0 \\0 &e^{-i\theta} \end{pmatrix}
\begin{pmatrix}
\phi_{dr} & \phi_{ur} \\ \widehat{\phi}_{ur}& \widehat{\phi}_{dr}
\end{pmatrix}
\end{align}
The equality (C.114) is equivalent to
\begin{align}
\begin{pmatrix}
\phi_{dg}' & \phi_{ug}' \\ \widehat{\phi}_{ug}'& \widehat{\phi}_{dg}'
\end{pmatrix}&=\begin{pmatrix} e^{-i\theta}&0 \\0 &e^{i\theta} \end{pmatrix}
\begin{pmatrix}
\phi_{dg} & \phi_{ug} \\ \widehat{\phi}_{ug}& \widehat{\phi}_{dg}
\end{pmatrix}
\end{align}
We get
\begin{align}
{\eta{'}}_{1dr}^*&=e^{-i\theta}\eta_{1dr}^*;~{\eta{'}}_{1ur}^*=e^{-i\theta}\eta_{1ur}^*\\
{\eta{'}}_{2dr}^*&=e^{-i\theta}\eta_{2dr}^*;~{\eta{'}}_{2ur}^*=e^{-i\theta}\eta_{2ur}^*\\
{\eta{'}}_{1dg}^*&=e^{i\theta}\eta_{1dg}^*;~{\eta{'}}_{1ug}^*=e^{i\theta}\eta_{1ug}^*\\
{\eta{'}}_{2dg}^*&=e^{i\theta}\eta_{2dg}^*;~{\eta{'}}_{2ug}^*=e^{i\theta}\eta_{2ug}^*
\end{align}
This gives
\begin{align}
s_1'&=s_1~;~~s_2'=e^{-i\theta}s_2~;~~s_3'=e^{i\theta}s_3 \\
s_4'&=s_4~;~~s_5'=e^{-i\theta}s_5~;~~s_6'=e^{i\theta}s_6 \\
s_9'&=s_9~;~~s_8'=e^{-2i\theta}s_8~;~~s_7'=e^{2i\theta}s_7 \\
s_{10}'&=s_{10}~;~~s_{11}'=e^{-i\theta}s_{11}~;~~s_{12}'=e^{i\theta}s_{12} \\
s_{14}'&=s_{14}~;~~s_{15}'=e^{-i\theta}s_{15}~;~~s_{13}'=e^{i\theta}s_{13} 
\end{align}
from which we get
\begin{align}
s_j'{s{'}}_j^*&=s_j s_j^*,~j=1,2,\dots,15 \\
\rho'&=\rho \\
\chi_r'&=e^{-\mathbf{i}\theta}\chi_r \\
\chi_g'&=e^{\mathbf{i}\theta}\chi_g
\end{align}
These relations are the awaited ones because
\begin{align}
\pmb\partial \Psi_r'&=\pmb\partial(e^{\mathbf{i}\theta}\Psi_r)\notag \\
&=e^{-\mathbf{i}\theta}(-\mathbf{i}\pmb\partial\theta\Psi_r +\pmb\partial\Psi_r)\\
\pmb\partial \Psi_g'&=\pmb\partial(e^{-\mathbf{i}\theta}\Psi_g)\notag \\
&=e^{\mathbf{i}\theta}( \mathbf{i}\pmb\partial\theta\Psi_g +\pmb\partial\Psi_g)\\
{\mathbf{G}{'}}^3 &=\mathbf{G}^3-\frac{2}{g_3}\pmb\partial\theta.
\end{align}

\subsection{Gauge group generated by $\underline{\mathbf{i}}\Gamma_8$}

We name $f_8$ the gauge transformation
\begin{equation}
f_8:\Psi^c \mapsto \underline{\mathbf{i}}\Gamma_8(\Psi^c)=\begin{pmatrix}
0 & \frac{\mathbf{i}}{\sqrt{3}} \Psi_r \\\frac{\mathbf{i}}{\sqrt{3}} \Psi_g & 
-\frac{2\mathbf{i}}{\sqrt{3}} \Psi_b
\end{pmatrix}
\end{equation}
which implies
\begin{align}
[\exp(\theta f_1)](\Psi^c)&=\begin{pmatrix}
0 & e^{\frac{\theta\mathbf{i}}{\sqrt{3}}}\Psi_r  \\ e^{\frac{\theta\mathbf{i}}{\sqrt{3}}}\Psi_g  
& e^{-\frac{2\theta\mathbf{i}}{\sqrt{3}}}\Psi_b
\end{pmatrix}=\begin{pmatrix}
0 & \Psi_r' \\ \Psi_g ' & \Psi_b '
\end{pmatrix}\\
\Psi_r'&=\exp(\frac{\theta\mathbf{i}}{\sqrt{3}})\Psi_r \\
\Psi_g '&=\exp(\frac{\theta\mathbf{i}}{\sqrt{3}})\Psi_g \\
\Psi_b '&= \exp(-\frac{2\theta\mathbf{i}}{\sqrt{3}})\Psi_b
\end{align}
This gives
\begin{align}
\phi_{dr}'&=\exp(\frac{i\theta}{\sqrt{3}})\phi_{dr};~\phi_{ur}'=\exp(\frac{i\theta}{\sqrt{3}})\phi_{ur}\\
\phi_{dg}'&=\exp(\frac{i\theta}{\sqrt{3}})\phi_{dg};~\phi_{ug}'=\exp(\frac{i\theta}{\sqrt{3}})\phi_{ug}\\
\phi_{db}'&=\exp(-\frac{2i\theta}{\sqrt{3}})\phi_{db};~\phi_{ub}'=\exp(-\frac{2i\theta}{\sqrt{3}})\phi_{ub}
\end{align}
We then get
\begin{align}
{\eta{'}}^*_{1dr}&=\exp(\frac{i\theta}{\sqrt{3}})\eta_{1dr}^*;~
{\eta{'}}^*_{1dg}=\exp(\frac{i\theta}{\sqrt{3}})\eta_{1dg}^*;~
{\eta{'}}^*_{1db}=\exp(-\frac{2i\theta}{\sqrt{3}})\eta_{1dg}^* \\
{\eta{'}}^*_{2dr}&=\exp(\frac{i\theta}{\sqrt{3}})\eta_{2dr}^*;~
{\eta{'}}^*_{2dg}=\exp(\frac{i\theta}{\sqrt{3}})\eta_{2dg}^*;~
{\eta{'}}^*_{2db}=\exp(-\frac{2i\theta}{\sqrt{3}})\eta_{2dg}^*\\
{\eta{'}}^*_{1ur}&=\exp(\frac{i\theta}{\sqrt{3}})\eta_{1ur}^*;~
{\eta{'}}^*_{1ug}=\exp(\frac{i\theta}{\sqrt{3}})\eta_{1ug}^*;~
{\eta{'}}^*_{1ub}=\exp(-\frac{2i\theta}{\sqrt{3}})\eta_{1ug}^* \\
{\eta{'}}^*_{2ur}&=\exp(\frac{i\theta}{\sqrt{3}})\eta_{2ur}^*;~
{\eta{'}}^*_{2ug}=\exp(\frac{i\theta}{\sqrt{3}})\eta_{2ug}^*;~
{\eta{'}}^*_{2ub}=\exp(-\frac{2i\theta}{\sqrt{3}})\eta_{2ug}^.
\end{align}
This implies
\begin{align}
s_1'&=\exp(\frac{2i\theta}{\sqrt{3}})s_1;~s_2'=\exp(-\frac{i\theta}{\sqrt{3}})s_2
;~s_3'=\exp(-\frac{i\theta}{\sqrt{3}})s_3 \\
s_4'&=\exp(\frac{2i\theta}{\sqrt{3}})s_4;~s_5'=\exp(-\frac{i\theta}{\sqrt{3}})s_5
;~s_6'=\exp(-\frac{i\theta}{\sqrt{3}})s_6 \\
s_7'&=\exp(\frac{2i\theta}{\sqrt{3}})s_7;~s_8'=\exp( \frac{2i\theta}{\sqrt{3}})s_8
;~s_9'=\exp(-\frac{4i\theta}{\sqrt{3}})s_9 \\
s_{10}'&=\exp(\frac{2i\theta}{\sqrt{3}})s_{10};~s_{11}'=\exp(-\frac{i\theta}{\sqrt{3}})s_{11}
;~s_{12}'=\exp(-\frac{i\theta}{\sqrt{3}})s_{12} \\
s_{13}'&=\exp(-\frac{i\theta}{\sqrt{3}})s_{13};~s_{14}'=\exp(\frac{2i\theta}{\sqrt{3}})s_{14}
;~s_{15}'=\exp(-\frac{i\theta}{\sqrt{3}})s_{15}. 
\end{align}
We then get the awaited results
\begin{align}
s_j'{s{'}}_j^*&=s_j s_j^*,~j=1,2,\dots,15 \\
\rho'&=\rho \\
\chi_r'&=\exp(-\frac{\mathbf{i}\theta}{\sqrt{3}})\chi_r \\
\chi_g'&=\exp(-\frac{\mathbf{i}\theta}{\sqrt{3}})\chi_g\\
\chi_b'&=\exp(\frac{2\mathbf{i}\theta}{\sqrt{3}})\chi_b .
\end{align}

\end{appendix}

\begin{thebibliography}{10}

\bibitem{davi:93}
C.~Daviau.
\newblock {\em Equation de Dirac non lin\'eaire}.
\newblock PhD thesis, Universit\'e de Nantes, 1993.

\bibitem{davi:97}
C.~Daviau.
\newblock Solutions of the {D}irac equation and of a nonlinear {D}irac equation
  for the hydrogen atom.
\newblock {\em Adv. Appl. Clifford Algebras}, 7((S)):175--194, 1997.

\bibitem{davi:05}
C.~Daviau.
\newblock Interpr\'etation cin\'ematique de l'onde de l'\'electron.
\newblock {\em Ann. Fond. L. de Broglie}, 30(3-4), 2005.

\bibitem{davi:11}
C.~Daviau.
\newblock {\em L'espace-temps double}.
\newblock JePublie, Pouill\'e-les-coteaux, 2011.

\bibitem{dav2:12}
C.~Daviau.
\newblock ${C}l_3^*$ invariance of the {D}irac equation and of
  electromagnetism.
\newblock {\em Adv. Appl. Clifford Algebras}, 22(3):611--623, 2012.

\bibitem{davi:12}
C.~Daviau.
\newblock {\em Double Space-Time and more}.
\newblock JePublie, Pouill\'e-les-coteaux, 2012.

\bibitem{dav:12}
C.~Daviau.
\newblock {\em Nonlinear Dirac Equation, Magnetic Monopoles and Double
  Space-Time}.
\newblock CISP, Cambridge UK, 2012.

\bibitem{dav:14}
C.~Daviau.
\newblock Gauge group of the standard model in ${C}l_{1,5}$.
\newblock {\em ICCA10, Tartu (Estonia)},
  http://hal.archives-ouvertes.fr/hal-01055145, 2014.

\bibitem{dabe:14}
C.~Daviau and J.~Bertrand.
\newblock Relativistic gauge invariant wave equation of the electron-neutrino.
\newblock {\em Journal of Modern Physics}, 5:1001--1022,
  http://dx.doi.org/10.4236/jmp.2014.511102, 2014.

\bibitem{davi:14}
C.~Daviau and J.~Bertrand.
\newblock {\em New Insights in the Standard Model of Quantum Physics in
  Clifford Algebra}.
\newblock Je Publie, Pouill\'e-les-coteaux, 2014 and
  http://hal.archives-ouvertes.fr/hal-00907848.

\bibitem{brog:24}
Louis de~Broglie.
\newblock Recherches sur la th\'eorie des quantas.
\newblock {\em Ann. Fond. Louis de Broglie}, 17(1), 1924.

\bibitem{dehe:93}
Ren\'e Deheuvels.
\newblock {\em Tenseurs et spineurs}.
\newblock PUF, Paris, 1993.

\bibitem{hest:86}
D.~Hestenes.
\newblock A unified language for {M}athematics and {P}hysics and {C}lifford
  {A}lgebra and the interpretation of quantum mechanics.
\newblock In Chisholm and AK~Common, editors, {\em Clifford Algebras and their
  applications in Mathematics and Physics}. Reidel, Dordrecht, 1986.

\bibitem{wein:67}
S.~Weinberg.
\newblock A model of leptons.
\newblock {\em Phys. Rev. Lett.}, 19:1264--1266, 1967.

\end{thebibliography}
\end{document}